\documentclass{PoS}
\usepackage{array,multirow}

\title{Status of (Direct and) Indirect Dark Matter searches}

\ShortTitle{Dark Matter}

\author{\speaker{Marco Cirelli}\thanks{I acknowledge the financial support of the European Research Council ({\sc Erc}) under the EU Seventh Framework Programme (FP7/2007-2013)/{\sc Erc} Starting Grant (agreement n.\ 278234 --- `{\sc NewDark}' project hosted by CNRS) and Advanced Grant (n.\ 267117 --- `{\sc Dark}' project hosted by Universit\'e Pierre \& Marie Curie - Paris 6.  I thank my collaborators and many of the participants / speakers in ICRC 2015 for very useful discussions. I acknowledge the hospitality of the Institut d'Astrophysique de Paris ({\sc Iap}) and the Public Library of the Comune di Milano (Italy) where part of this work was done.}\\
        IPhT, Universit\'e Paris-Saclay, CNRS, CEA, F-91191 Gif-sur-Yvette, France \& \\
        LPTHE, CNRS, UMR 7589, 4 Place Jussieu, F-75252, Paris, France \\
        E-mail: \email{marco.cirelli@gmail.com}}

\abstract{This article summarizes the status of the Indirect and Direct searches for Dark Matter, with a special focus and a critical look on the former. 
It is the write-up of a rapporteur talk given at the 34th International Cosmic Ray Conference (ICRC 2015) in The Hague, The Netherlands.}

\FullConference{Proceedings of the Corfu Summer Institute 2015 "School and Workshops on Elementary Particle Physics and Gravity"\\
		1-27 September 2015\\
		Corfu, Greece}

\begin{document}

\section{Introduction}
\label{sec:intro}

Cosmology and astrophysics provide several convincing evidences of the existence of Dark Matter (DM). 
The observation that some mass is missing to explain the internal dynamics of galaxy clusters and the rotations of galaxies dates back respectively to the '30s and the '70s. The observations from weak lensing, for instance in the spectacular case of the so-called `bullet cluster', provide evidence that there is mass where nothing is optically seen. More generally, global fits to a number of cosmological datasets (Cosmic Microwave Background, Large Scale Structure and also Type Ia Supernovae) allow to determine very precisely the amount of DM in the global energy-matter content of the Universe at $\Omega_{\rm DM} h^2 =0.1188 \pm 0.0010$~\cite{Ade:2015xua}\footnote{Here $\Omega_{\rm DM} = \rho_{\rm DM}/\rho_c$ is defined as usual as the energy density in Dark Matter with respect to the critical energy density of the Universe $\rho_c = 3 H_0^2/8\pi G_N$, where $H_0$ is the present Hubble parameter. $h$ is its reduced value $h = H_0 / 100\ {\rm km}\, {\rm s}^{-1} {\rm Mpc}^{-1}$.}.
 
All these signals pertain to the gravitational effects of Dark Matter at the cosmological and extragalactical scale. Searches for explicit manifestation of the DM particles that are supposed to constitute the halo of our own galaxy (and the large scale structures beyond it) have instead so far been giving negative results, but this might be on the point of changing. 

\smallskip

Indirect searches for DM are of particular interest for the cosmic ray community gathered at the 2015 edition of the International Cosmic Ray Conference (ICRC). These searches aim at detecting the signatures of the annihilations or decays of DM particles in the fluxes of Cosmic Rays (CRs), intended in a broad sense: charged particles (electrons and positrons, antiprotons, antideuterium), photons (gamma rays, X-rays, synchrotron radiation), neutrinos. 
In general, a key point of all these searches is to look for channels and ranges of energy where it is possible to beat the `background' from ordinary astrophysical processes (which are, ironically, the usual study matter of CR physicists!). This is for instance the basic reason why searches for charged particles focus on fluxes of antiparticles (positrons, antiprotons, antideuterons), much less abundant in the Universe than the corresponding particles, and searches for photons or neutrinos have to look at areas where the DM-signal to astro-noise ratio can be maximized. 
Pioneering works have explored indirect detection (ID) as a promising avenue of discovery since the late-70's. Since then, innumerable papers have explored the predicted signatures of countless particle physics DM models. 

\smallskip
DM can also be looked for via Direct Detection (DD), which will be briefly addressed in Sec.~\ref{sec:DD}. 
Finally, DM particles can be searched for by trying to produce them at particle colliders, such as the LHC at CERN. This strategy, which is definitely promising and in full swing now, has however essentially not been discussed at ICRC 2015, so I will not develop it further here.

\smallskip

Before moving to the subject matter, let us quickly remind ourselves of the framework inside which most of the current activity develops: the one centered around WIMPs. 
A well spread theoretical prejudice wants the DM particles to be thermal relics from the Early Universe. They were as abundant as photons in the beginning, being freely created and destructed in pairs when the temperature of the hot plasma was larger then their mass. Their relative number density started then being suppressed as annihilations proceeded but the temperature dropped below their mass, due to the cooling of the Universe. Finally the annihilation processes also froze out as the Universe expanded further. The remaining, diluted abundance of stable particles constitutes the DM today. As it turns out, particles with weak scale mass ($\sim 100\, {\rm GeV} - 1\, {\rm TeV}$) and weak interactions could play the above story remarkably well, and their final abundance would automatically (miracolously?) be the observed $\Omega_{\rm DM}$. While this is certainly not the only possibility, the mechanism is appealing enough that a several-GeV-to-some-TeV scale DM particle with weak interactions (WIMP) is often considered as the most likely DM candidate.
An important corollary of the long-term fascination of the community for the WIMP miracle, or more generally the thermal relic production mechanism, is that DM particles are expected to annihilate in pairs into Standard Model particles. More precisely, a velocity averaged annihilation cross section of $\langle \sigma v \rangle = 3 \cdot 10^{-26} \ {\rm cm}^3/{\rm s}$ is seen as the benchmark value, since it is the one that yields the correct relic abundance. Such annihilations, occurring today in the Milky Way, could hence originate the anomalous CRs looked for in Indirect Detection. It is also possible, however, that DM particles decay: provided that their half-life is long enough that such process decay does not deplete the cosmological density significantly (otherwise it would contradict the experimental evidences for DM existence), the decay process could then also be at the origin of CRs to be searched for in ID. 

In any case, even independently of the theory prejudices, the mass range around TeV-ish DM is being explored right now and will certainly be the focus of even more intense explorations in the near future, by the ID, the DD and the collider experiments.

\section{Indirect detection via charged cosmic rays}
\label{sec:chargedCR}

\subsection{Electrons and positrons}
\label{sec:epem}

\begin{figure}[t]
\begin{center}
\includegraphics[width=0.48 \columnwidth]{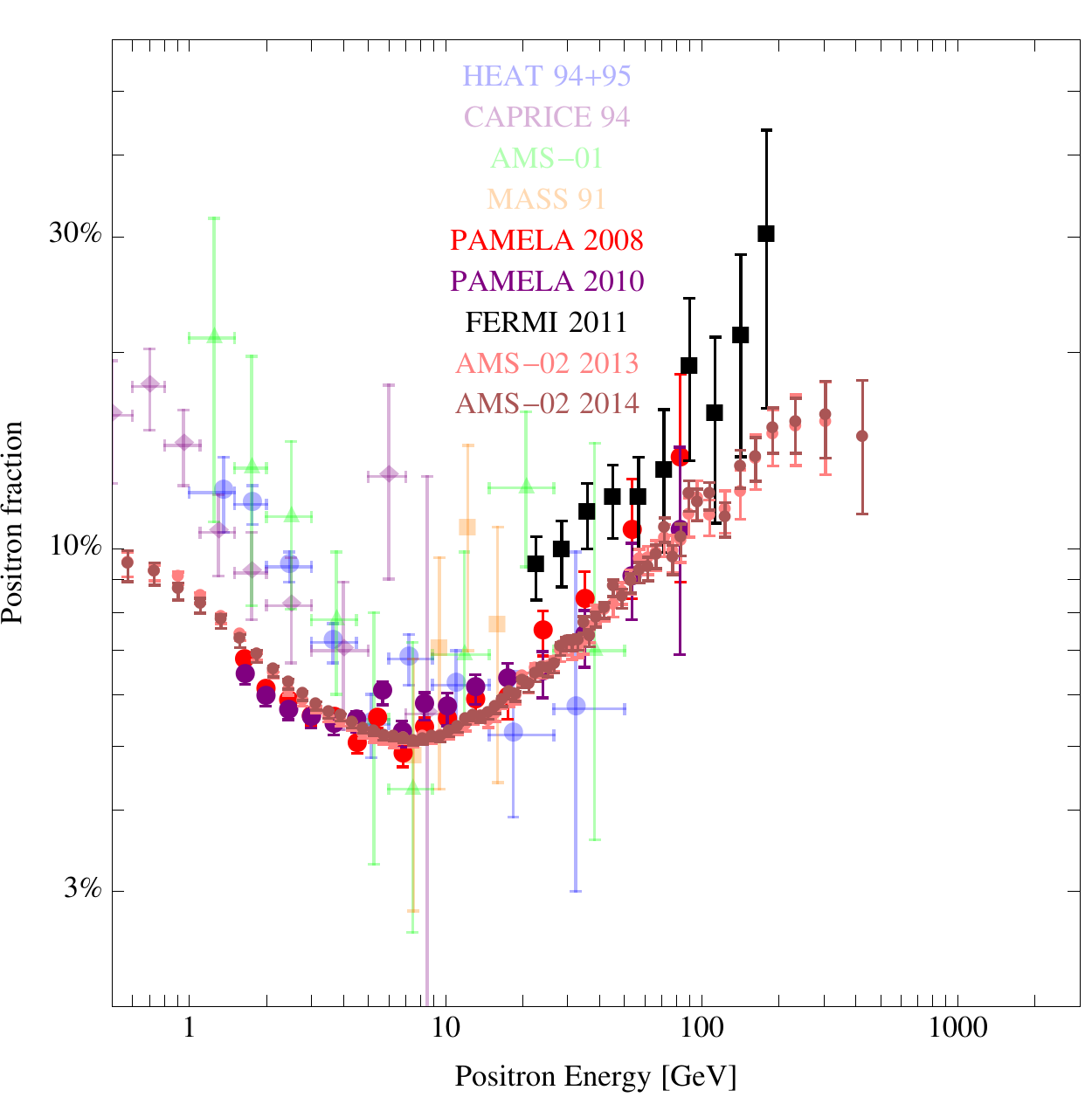} \
\includegraphics[width=0.497 \columnwidth]{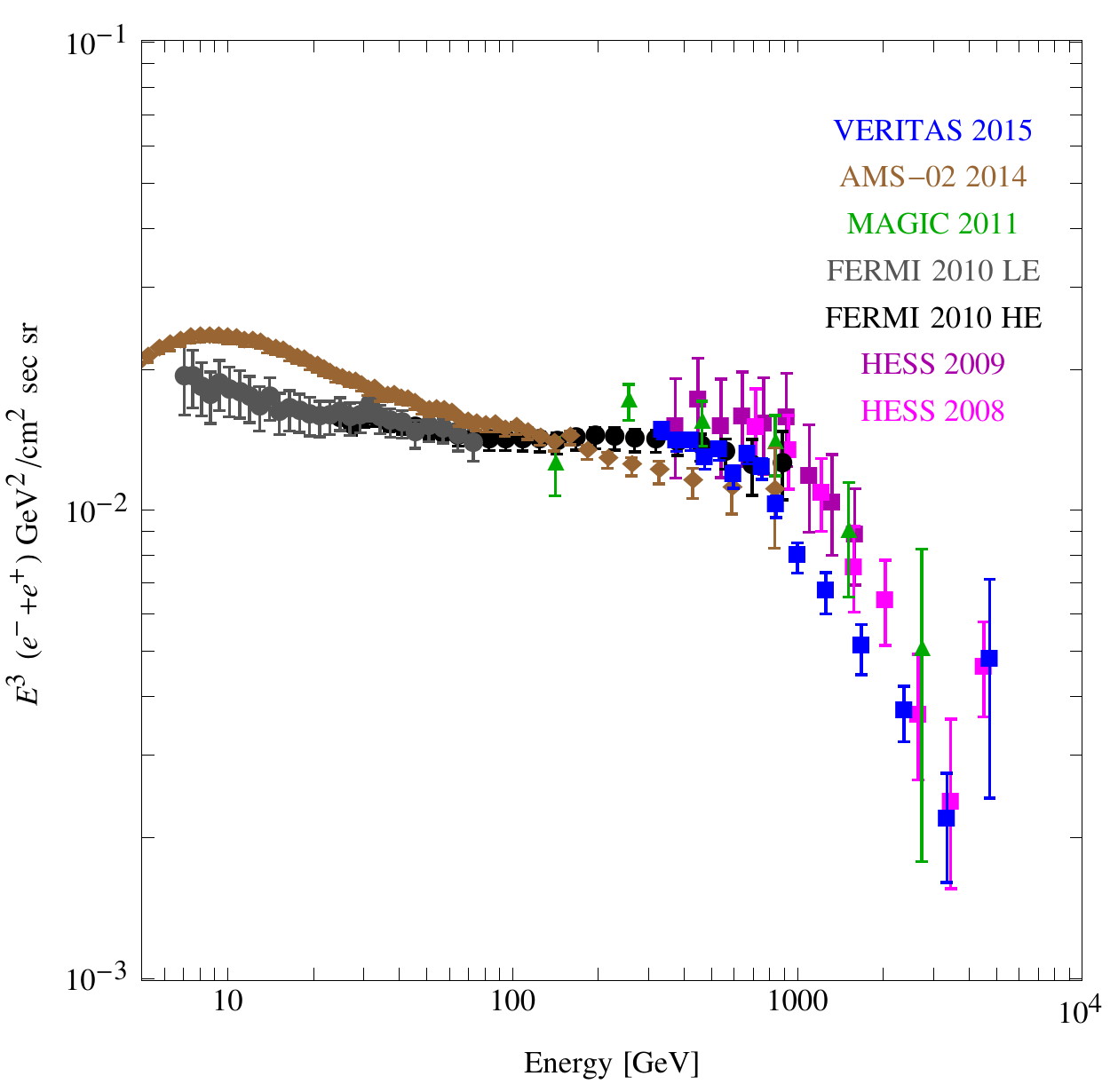}
\caption{A compilation of recent and less recent data in charged cosmic rays. Left: positron fraction. Right: sum of electrons and positrons.}
\label{fig:chargeddata}
\end{center}
\end{figure}

Since 2008 or so, there has been a flurry of positive results from a few indirect detection experiments looking at the fluxes of electrons and positrons, pointing in particular to `excesses' at the TeV and sub-TeV scale. A selection of these results is collected in fig.~\ref{fig:chargeddata}. 

Notorious data from the {\sc Pamela} satellite~\cite{PAMELApositrons,boezio} showed a steep increase in the energy spectrum of the positron fraction $e^+/(e^++e^-)$ above 10~GeV up to 100~GeV, compatibly with previous hints from {\sc Heat}~\cite{HEAT} and {\sc Ams-01}~\cite{AMS-01}. Qualitatively, these findings have been confirmed, and extended to about 200 GeV, with an independent measurement by the {\sc Fermi} satellite~\cite{FERMIpos}, although the normalization differs somewhat. More recently, in 2013 and 2014, the {\sc Ams-02} experiment on board the International Space Station has produced high accuracy data~\cite{ting,kounine,Aguilar:2013qda,Accardo:2014lma} that have allowed to confirm the {\sc Pamela} $e^+$ rise and also to point out a possible flattening above 300 GeV. 

In the measurement of the sum of electrons and positrons $(e^++e^-)$, data up to 1 TeV are provided by the {\sc Fermi} satellite~\cite{FERMIleptons} as well as, more recently, by {\sc Ams-02}~\cite{ting,duranti,Aguilar:2014fea}: they both indicate a rather featureless spectrum, although the spectral indexes differ, the one of {\sc Ams-02} being steeper. This mild disagreement will probably be settled soon, since the {\sc Fermi} collaboration has identified a systematic effect which will likely bring its data closer to those of {\sc Ams-02} in the upcoming Pass-8 release~\cite{FERMIvsAMS}. Above 1 TeV, data are provided by the {\sc Hess}~\cite{HESSleptons} and {\sc Magic}~\cite{BorlaTridon:2011dk} telescopes and show some indication for a cutoff or a steepening at energies of a few TeV. At ICRC 2015, the {\sc Veritas} telescope has presented new data~\cite{staszak} which also go in the same direction. 

\smallskip

The signals presented above are therefore striking because they imply the existence of a source of {\em `primary'} $e^+$ (and $e^-$) other than the ordinary astrophysical ones. This unknown new source can well be itself of astrophysical nature, e.g. one or more pulsar(s) / pulsar wind nebula(\ae), supernova remnants etc: this possibility is discussed in detail in several contributions to these ICRC 2015 proceedings~\cite{serpico,boudaud,dimauro,grimani}. 
It is however very tempting to try and read in these `excesses' the signature of DM. 

Indeed, as already mentioned above, the DM particles that constitute the DM halo of the Milky Way are expected to annihilate (or perhaps decay) into pairs of primary SM particles (such as $b \bar b$, $\mu^+\mu^-$, $\tau^+\tau^-$, $W^+W^-$ and so on) which, after decaying and through the processes of showering and hadronizing, give origin to fluxes of energetic cosmic rays: $e^-, e^+, \bar p$ (and also $\gamma$-rays, $\nu$...). Depending on which one has been the primary SM particle, the resulting spectra differ substantially in the details. Generically, however, they feature a `bump'-like shape, characterized by a high-energy cutoff at the DM particle mass and, for $e^\pm$ in particular, a softly decreasing tail at lower energies. It is thus very natural to expect a DM source to `kick in' on top of the secondary background and explain the $e^\pm$ excesses. The energy range, in particular, is tantalizingly right: the theoretically preferred TeV-ish DM would naturally give origin to TeV and sub-TeV bumps and rises.

The $e^-$, $e^+$ and $\bar p$ produced in any given point of the halo propagate immersed in the turbulent galactic magnetic field. This is exactly analogous to what ordinary charged cosmic rays do (with the only difference that ordinary CRs are mainly produced in the disk). The field consists of random inhomogeneities that act as scattering centers for charged particles, so that their journey can effectively be described as a diffusion process from an extended source (the DM halo) to some final given point (the location of the Earth, in the case of interest). While diffusing, charged CRs experience several other processes, and in particular energy losses due to synchrotron radiation, Inverse Compton Scattering (ICS) on the low energy photons of the CMB and starlight, Coulomb losses, bremsstrahlung, nuclear spallations...\,.
The transport process is solved numerically or semi-analytically using codes such as {\sc Galprop}~\cite{galprop}, {\sc Dragon}~\cite{Gaggero:2015mga}, {\sc Usine}~\cite{usine}, {\sc Picard}~\cite{picard}.

\begin{figure}[t]
\begin{center}
\includegraphics[width=0.44 \columnwidth]{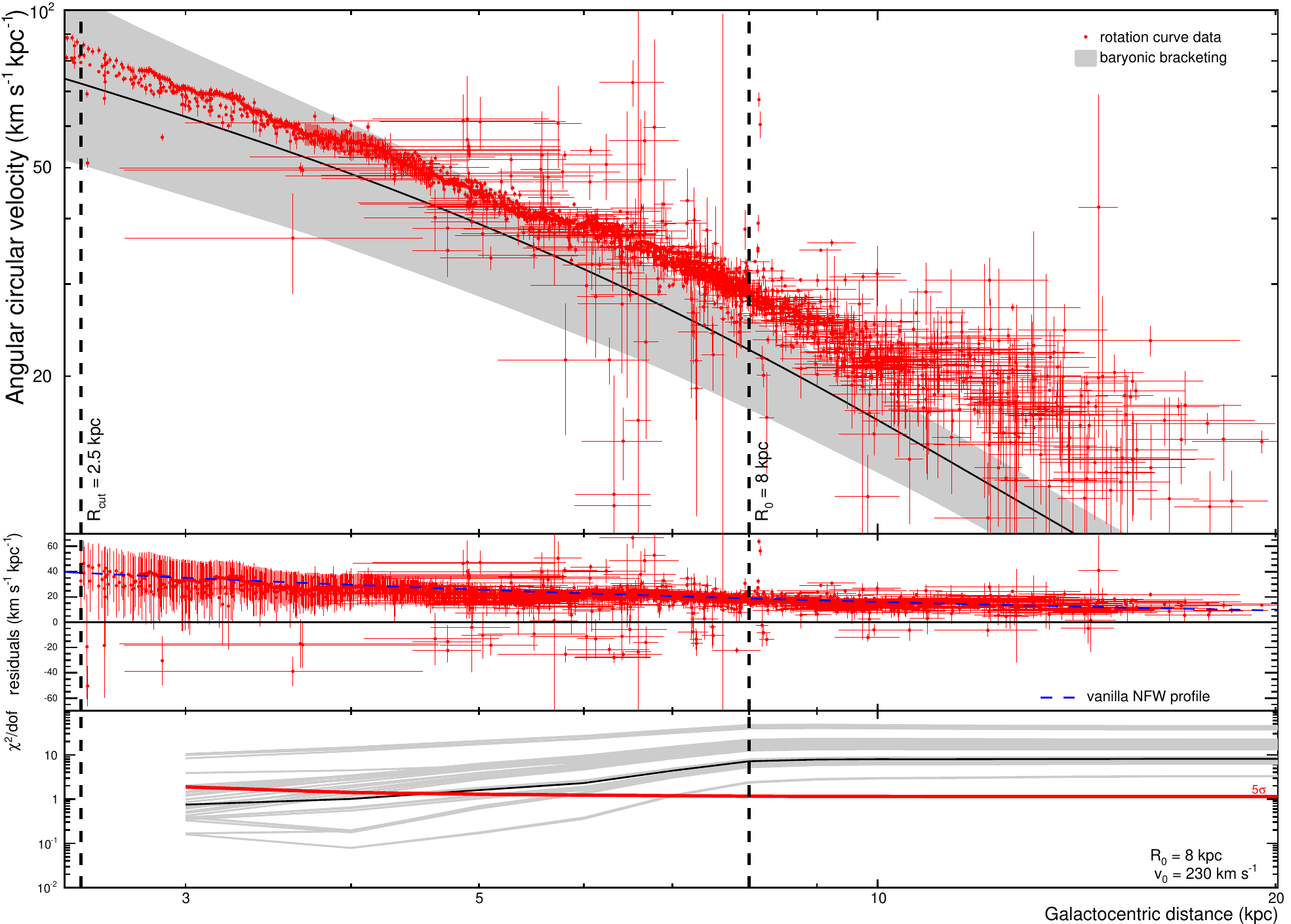} \
\includegraphics[width=0.54 \columnwidth]{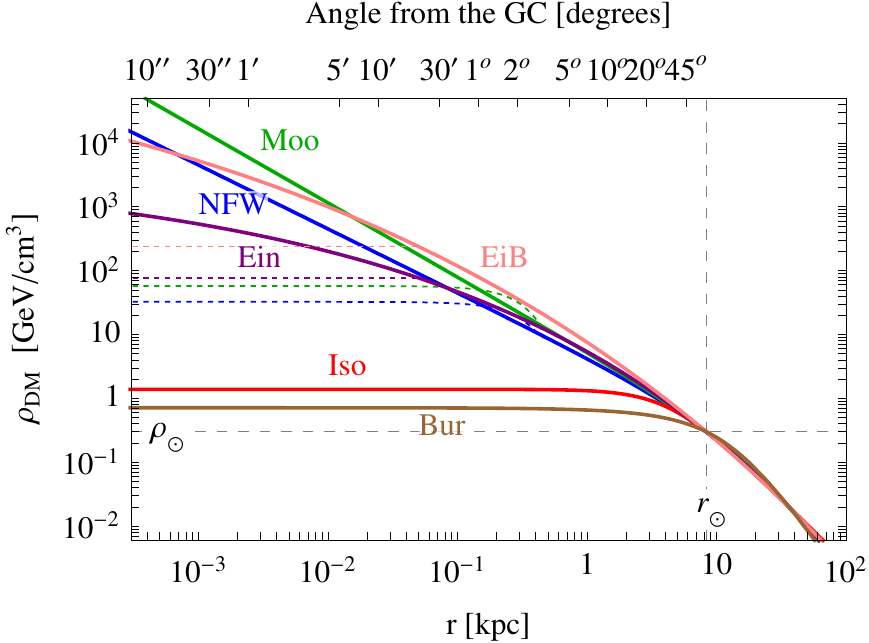}
\caption{Left: a collection of galactic kinematical data and their use to claim evidence for DM in the Galaxy as well as, in principle, to determine its distribution (figure from~\cite{pato,Iocco:2015xga}). Right: an illustration of the most-often used DM profiles (from~\cite{Cirelli:2010xx}).}
\label{fig:DMdistribution}
\end{center}
\end{figure} 

The source, DM annihilations or decays, follows $\rho(\vec x)$, the DM density distribution in the galactic halo, to the first power (in case of decays) or to the second power (for annihilations). What to adopt for $\rho(\vec x)$ is another one of the main open problems in the field, which has been discussed at some length at ICRC 2015~\cite{pato,Iocco:2015xga,silverwood}. 
Based on the results of increasingly more refined numerical simulations or on direct observations~\cite{pato,Iocco:2015xga}, profiles that differ even by several orders of magnitude at the GC are routinely adopted: e.g. the classical Navarro-Frenk-White (NFW) or the Einasto one, which exhibit a cusp at the galactic center, or the truncated isothermal or the Burkert one, which feature a central core. All profiles, on the other hand, are roughly normalized at the same value at the location of the Earth, which needs to be determined accurately~\cite{silverwood}. 
These features generically imply that observables which depend mostly on the local DM density (for instance, the flux of high energy positrons, which cannot come from far away due to energy losses) will not be very affected by the choice of profile, while those that are sensitive to the density at the GC will be affected the most (e.g. gamma rays observations of regions close to the GC).

\begin{figure}[t]
\begin{center}
\includegraphics[width=0.58 \columnwidth]{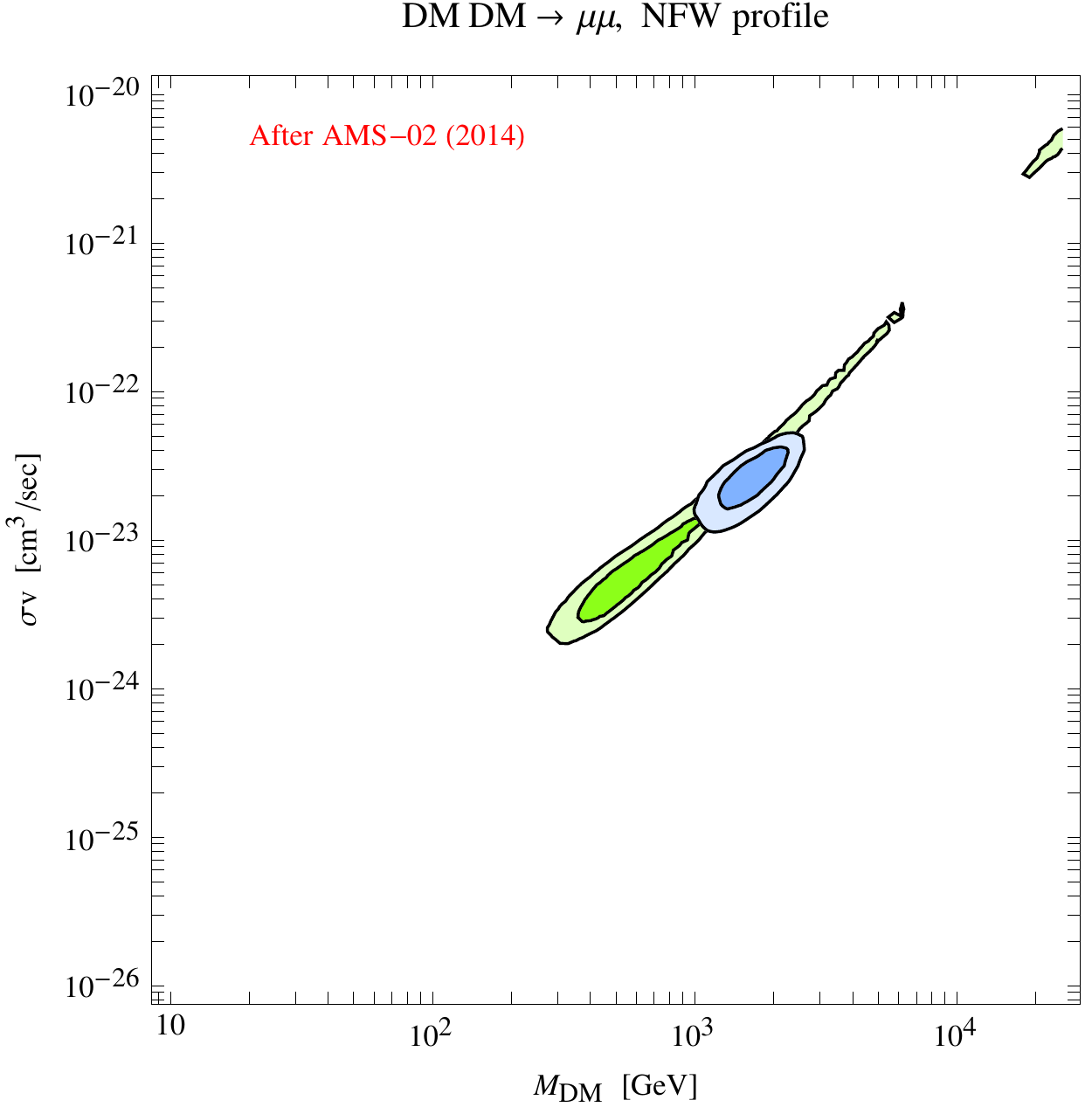}
\caption{An example of the current status of the global fits of positron fraction data (green areas) and of electron+positron data (blue regions), for one specific case of DM profile and annihilation channel. Figure produced for ICRC 2015.}
\label{fig:chargedfit}
\end{center}
\end{figure} 

\begin{figure}[t]
\begin{center}
\includegraphics[width=0.45 \columnwidth]{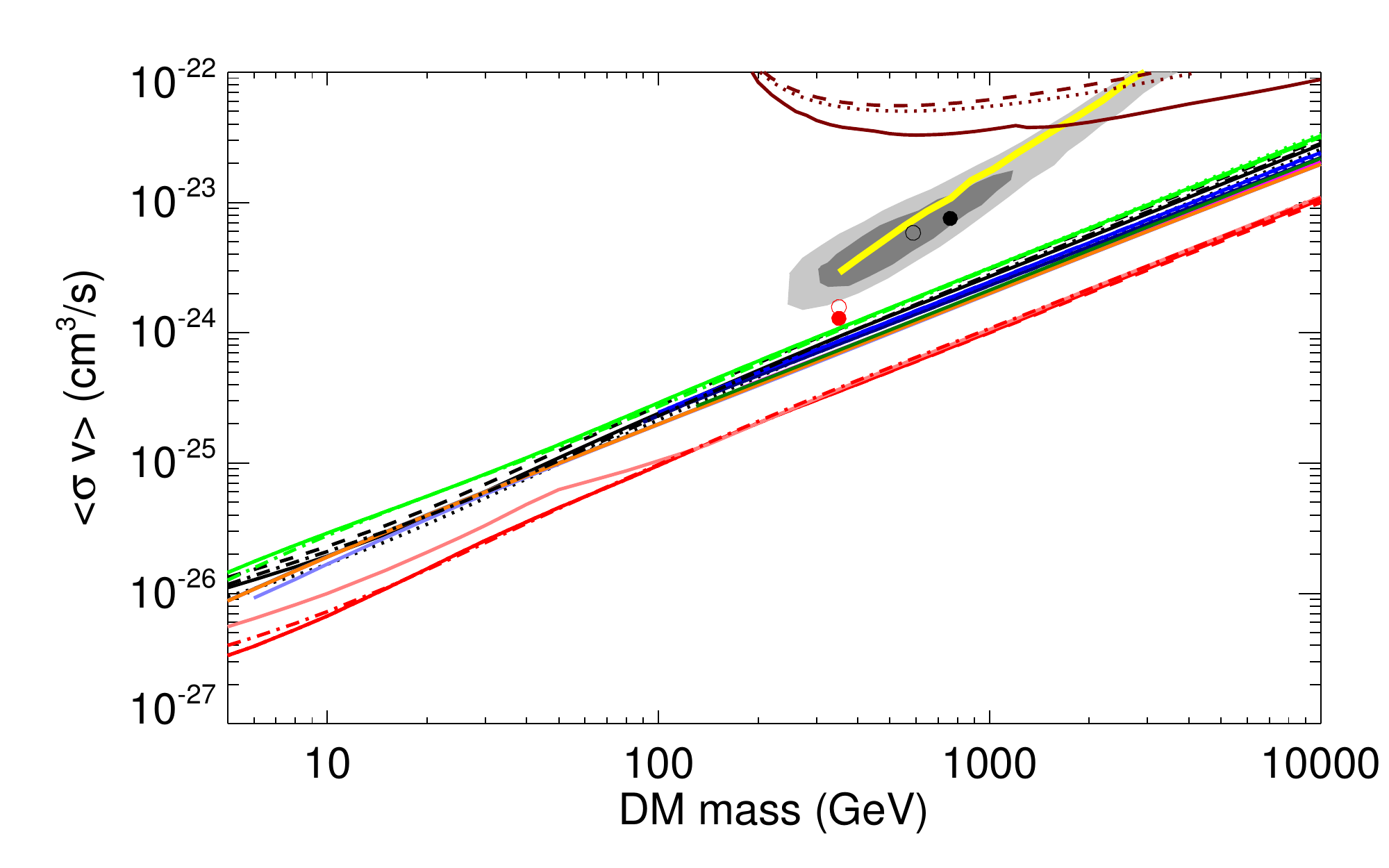} \
\includegraphics[width=0.45 \columnwidth]{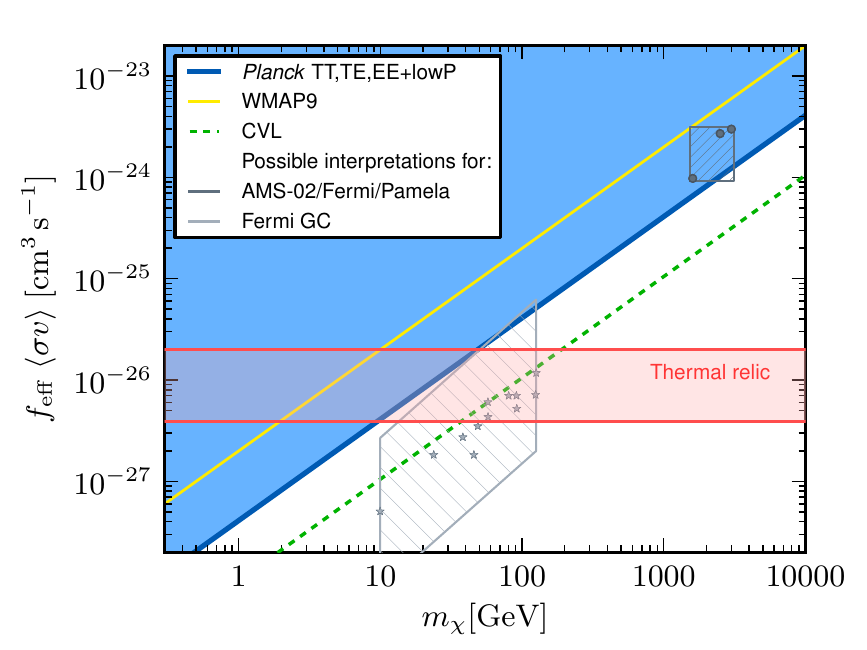}
\caption{CMB bounds on DM annihilation (figures from~\cite{Slatyer:2015jla} and~\cite{Ade:2015xua}).}
\label{fig:CMBbounds}
\end{center}
\end{figure}

All in all, the DM annihilation (or decay) channel, the DM mass and the annihilation cross section (or decay rate), convoluted with the information on the propagation process and the DM galactic profile allow to determine the expected CR fluxes and compare them with the data. Of course, a proper treatment of the background from astrophysics is crucial to obtain meaningful results (this step often represents the most tricky one in the actual analysis). 
Under conservative and rather broad assumptions for the background and for propagation, an example of the results of global fits to data, for the case of annihilations, is given in fig.~\ref{fig:chargedfit}. As perhaps well known, the DM needed to fit the data has to: have a mass in the TeV / multi-TeV range, have a very large annihilation cross section (of the order of $10^{-23} {\rm cm}^3/s$, orders of magnitude larger than the cosmological prejudice) and be leptophilic (to avoid contradicting antiproton bounds for such a large cross section, see the next section). So a global Dark Matter interpretation of the leptonic `excesses' can be attempted. However, even restricting to leptonic data only, some tension is present. First, as apparent in fig.~\ref{fig:chargedfit}, the green and blue regions overlap only marginally, i.e. the positron data tend to prefer a mass smaller than the all lepton ones. This is due to the bending of the highest energy {\sc Ams-02} data~\cite{Cirelli:2008pk}. Secondly, the conclusions reached above may be questioned if one fully exploits the precision of the {\sc Ams-02} data and if one relaxes the assumptions (e.g.~considering more annihilation channels with different branching ratios), as done in~\cite{boudaud}. Finally, and perhaps most importantly, a significant tension exists with constraints from gamma rays and from the CMB. We will discuss the former ones in a subsequent section. The CMB ones stem from the fact that DM annihilations in the Early Universe inject energy that modifies the properties of the CMB, mainly via the induction of excessive ionized material at early redshift. Fig.~\ref{fig:CMBbounds} reports two recent determinations of the bounds, that manifestly exclude the DM interpretation of the leptonic `excesses'~\cite{Ade:2015xua,Slatyer:2015jla}. These constraints have the advantage of being insensitive to the usual astrophysical uncertainties that affect the gamma and charged CR ray bounds (e.g. the DM profile), but they can be evaded if the cross section is suppressed at low velocities or early times.

\smallskip

So keeping the DM option on the table, the question becomes: how will we be able to discriminate between a DM versus an astrophysical origin of the excesses? It is repeatedly claimed, e.g.~by the press releases of the {\sc Ams-02} collaboration~\cite{ting,AMSpress1}, that the behavior of the positron fraction spectrum at high energies, which will be precisely measured by the experiment, will allow such discrimination: the assumption is that a DM spectrum would show a sharp cut-off (at the value corresponding to the particle's mass) while an astrophysical source would present a smoother decline. While it is certainly true that a very important role will be played by precision measurements, unfortunately the above statements are too simplistic. Indeed, for instance, when DM annihilates in a combination of channels, including hadronic ones, its $e^+$ spectrum can be soft. On the other hand, a sufficiently young and nearby pulsar could explain the data and would produce a steep spectral descent. In general, too little is known on the details of CR propagation and on the mechanisms of local astrophysical sources to support the simplistic claims on the shape of the fraction.

A more promising avenue is connected to the arrival direction of the cosmic leptons. Detecting a clear dipole anisotropy would be a clear indication of the existence of a single point source while measuring an isotropic distribution would rather favor a diffuse source such as DM (or a collection of uniformly distributed point sources). The anisotropy measurement is most useful if performed as a function of the energy of the incoming leptons: indeed, low energy cosmic rays are generically predicted to come from farther away (a long propagation degrades the energy) and therefore to have a randomized arrival direction, while high energy ones would better retain the information. {\sc Fermi} has reported upper bounds as a function of the energy in~\cite{Ackermann:2010ip}. {\sc Ams-02} has reported a flat upper bound of 3\% in~\cite{Accardo:2014lma} and plans to reach a 1\% sensitivity~\cite{ting} in the dipole anisotropy. {\sc Pamela} has recently quoted an upper bound of 16\%~\cite{Adriani:2015kfa}.

\smallskip

Very fortunately, the near future has in store more precision data on leptons. At ICRC 2015, the japanese-led {\sc Calet} detector~\cite{calet} and the chinese-led {\sc Dampe} satellite~\cite{dampe} were presented in some detail: they will explore an energy range up to several TeV with unprecedented resolution. {\sc Calet} has been installed on the ISS in august 2015 while {\sc Dampe} is scheduled for launch in december 2015. Remarkably enough, even the future gamma ray imaging telescope {\sc Cta} plans to be able to measure the positron fraction at high energies, using the Earth-Moon system as a spectrometer~\cite{CTApositrons}.

\subsection{Antiprotons}
\label{sec:antiprotons}

\begin{figure}[t]
\includegraphics[width=0.56 \columnwidth]{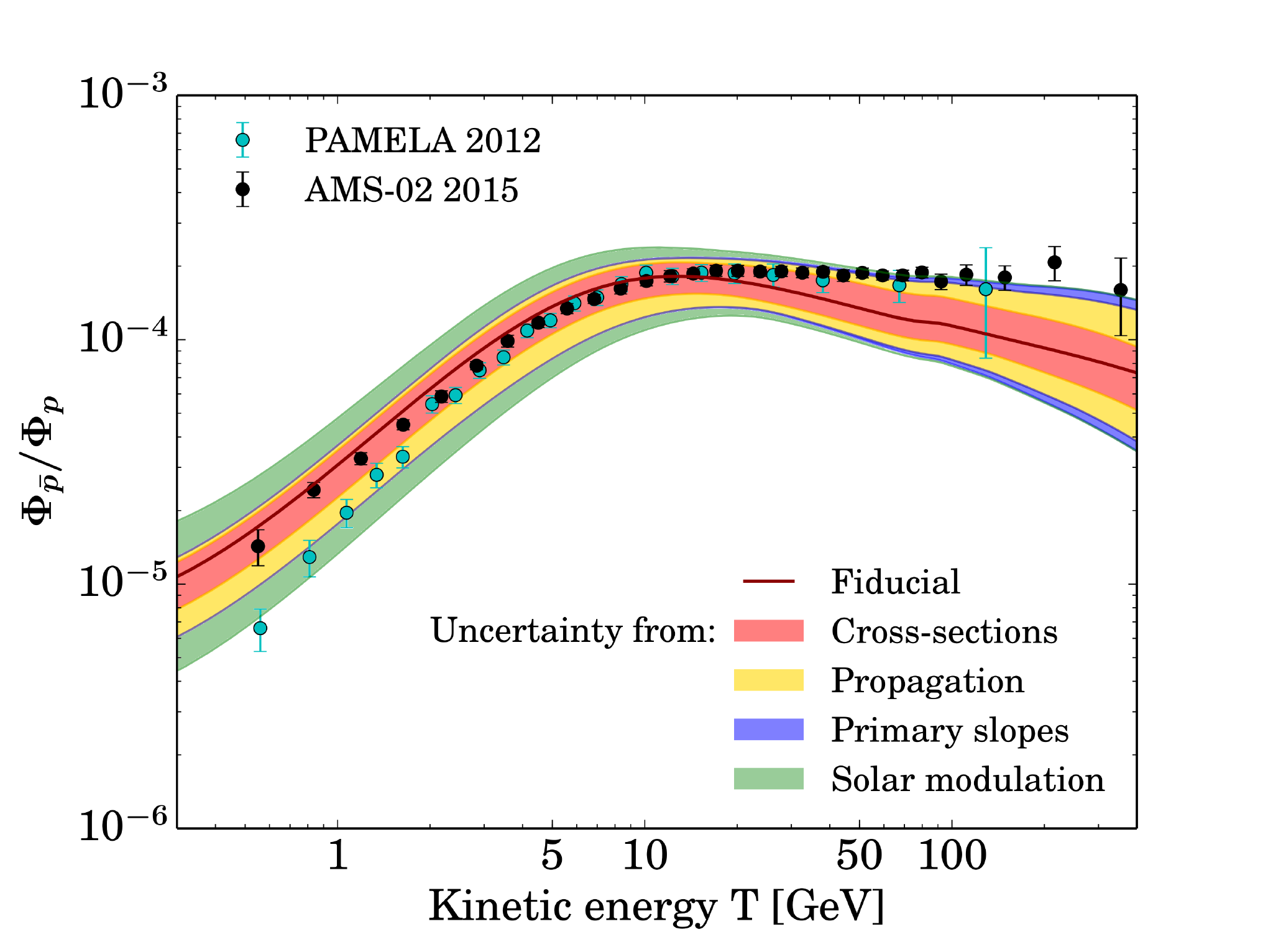} \hspace{-0.8cm}
\includegraphics[width=0.42 \columnwidth]{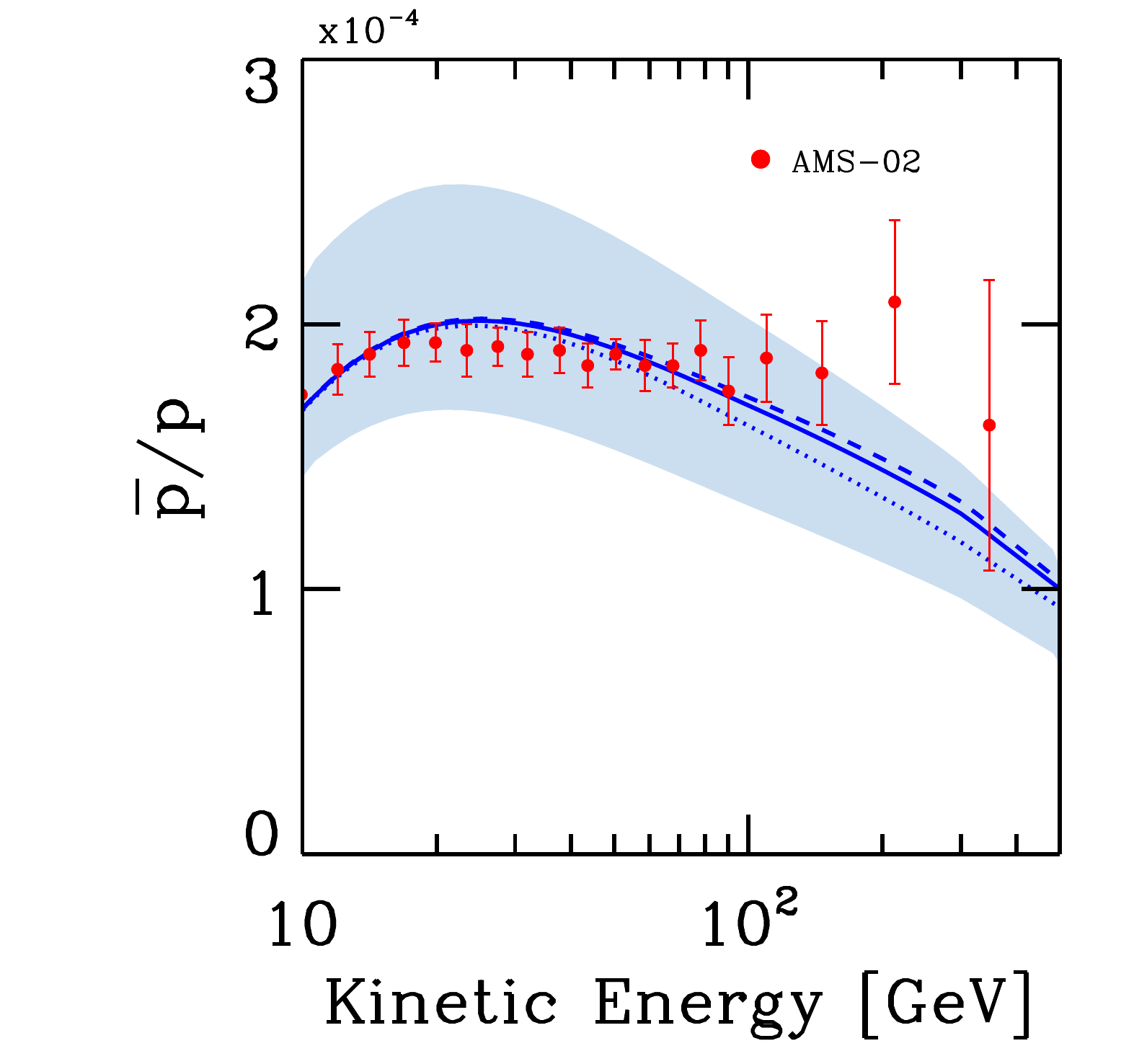} \\
\includegraphics[width= 0.58 \columnwidth]{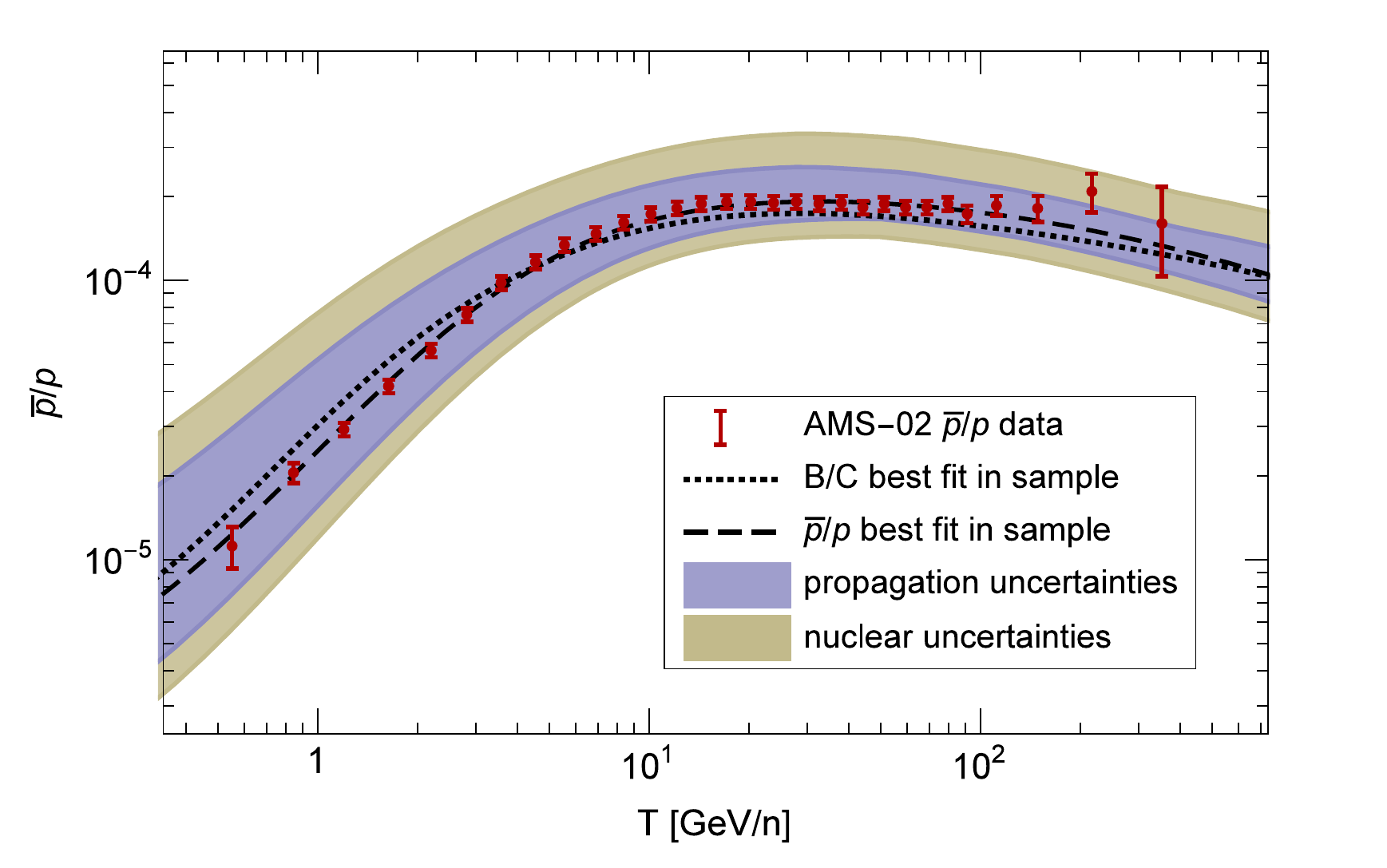}
\includegraphics[width= 0.4 \columnwidth]{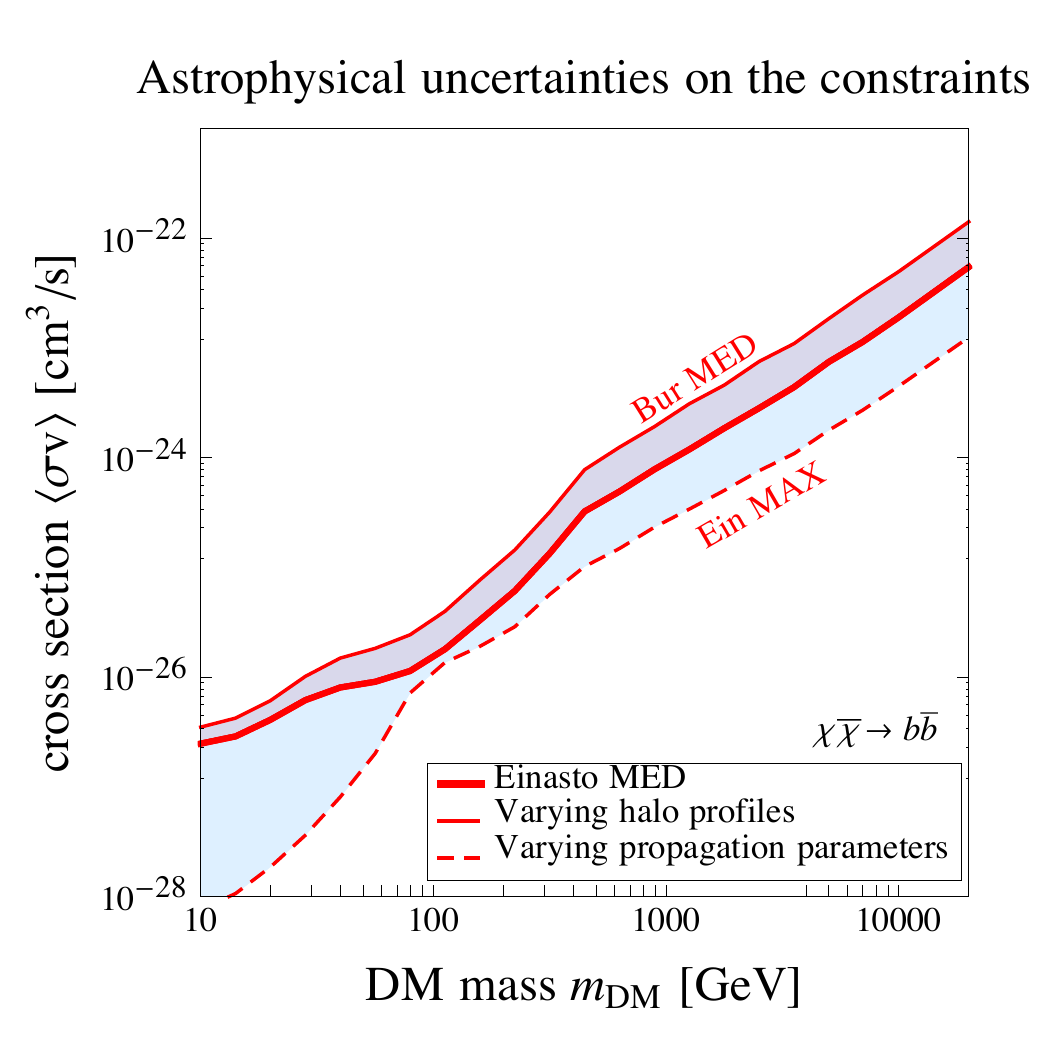}
\caption{Antiproton measurements by {\sc Pamela} and {\sc Ams-02} compared to the astrophysical prediction and its uncertainties, in three independent analyses, and the constraints on DM annihilation that originate from them. Top-left and bottom-right figures are from~\cite{Giesen:2015ufa}; top-right from~\cite{Evoli:2015vaa}, bottom-left from~\cite{Kappl:2015bqa}.}
\label{fig:antiprotons}
\end{figure} 

In the antiproton channel, data have been published by {\sc Pamela} since 2008~\cite{Adriani:2008zq} (and then 2010~\cite{Adriani:2010rc} and 2012~\cite{Adriani:2012paa}). More recently, {\sc Ams-02} has also presented {\em preliminary} data on the $\bar p/p$ ratio, in a a widely advertised event at CERN~\cite{amsdays} and then at ICRC 2015~\cite{ting,kounine}. The data-sets from the two experiments, reported in fig.~\ref{fig:antiprotons}, are in very good mutual agreement, although the {\sc Ams-02} ones are of course much more accurate and extend to higher energies. 

In the {\sc Ams} presentation strategy~\cite{ting,kounine,pressrelease}, the public may be led to believe that the data are at odds with the predictions from astrophysics and therefore that a new component (Dark Matter!) has to be invoked. However, despite the extent to which we would all love {\sc Ams} to find something extraordinarily new, this is at best clearly premature. Indeed, when one includes, in the computations of the predictions from astrophysics, the latest recent developments, the discrepancy is largely reabsorbed. Such latest developments include: (i) the new measurement of the primary proton and Helium spectra (which, impinging on the interstellar medium, produce the bulk of the astrophysical antiprotons), as delivered by {\sc Ams} itself~\cite{choutko,Aguilar:2015ooa}~\footnote{These measurements are now in very good agreement with the corresponding measurements by {\sc Pamela} from 2011~\cite{Adriani:2011cu,boezio}, after that {\sc Ams} has revisited its understanding of the systematic errors with respect to the data presented at ICRC 2013 (see footnote 23 of \cite{choutko}): now both experiments see a spectral break (a.k.a. a `change of slope') at about 300 GeV.}; (ii) the recent results on the antiproton spallation production cross section~\cite{diMauro:2014zea,moskalenko}; (iii) updated propagation schemes~\cite{genolini}\ldots 

A collection of fairer comparisons between the data and the updated predictions from astrophysics is reported in fig.~\ref{fig:antiprotons} (first 3 panels). The predictions differ slightly because they employ different inputs and different determinations of the uncertainties. However, the qualitative conclusion is quite apparent: contrarily to the leptonic case, there is no unambiguous excess in antiproton data. On the other hand, it is true that the data seem to prefer a rather flat $\bar p/p$, which is somewhat difficult to obtain in current astrophysical models. This may point to propagation schemes characterized by a relatively mild energy dependence of the diffusion coefficient at high energies. Although it is too early to draw strong conclusions, this is an interesting observation and it goes in the same direction as the preference displayed by the preliminary B/C {\sc Ams-02} data.

For what concerns Dark Matter, since no excess can undisputedly be claimed, one can derive constraints~\cite{Giesen:2015ufa,boudaudpbar}. This is what is reported in the last panel of fig.~\ref{fig:antiprotons} for one specific example. Fixing a benchmark DM profile (Einasto) and the {\sc Med} propagation scheme, the constraints exclude the thermal annihilation cross section $\langle \sigma v \rangle = 3 \cdot 10^{-26} \ {\rm cm}^3/{\rm s}$ for $m_{\rm DM} \sim 150$ GeV. The modification of the profile or the propagation scheme has the effect of spanning the shaded band, i.e.~affecting the bounds by a factor of a few. For analogous constraints on decaying DM, see~\cite{Giesen:2015ufa,boudaudpbar,Mambrini:2015sia}.

\subsection{Antideuterium}
\label{sec:antideuterium}

\begin{figure}[t]
	\parbox[b]{.45\linewidth}{
		\includegraphics[width=\linewidth]{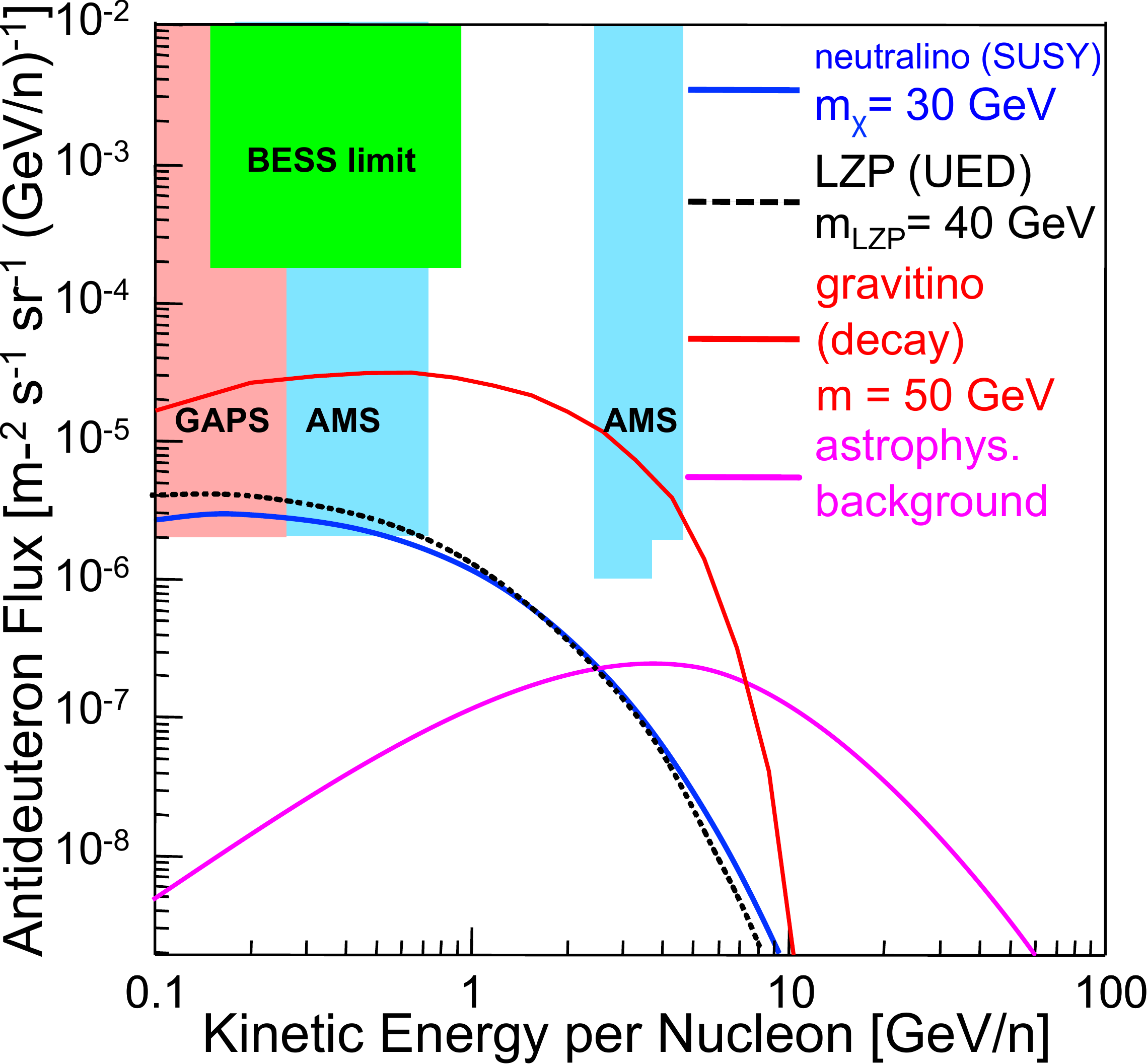}}\hfill
	\parbox[b]{.45\linewidth}{
		\caption{Spectra of antideuterons as predicted by some benchmark Dark Matter models (solid lines) superimposed to the predicted background from astrophysics, the current {\sc Bess} limit (green shaed area) and the projected sensitivity of {\sc Gaps} (pink shaded area) and {\sc Ams} (blue shaded area). The figure is taken from~\cite{vonDoetinchem:2015zva}. }
		\label{fig:antideuteron}
	}
\end{figure} 

Antideuterons (the bound states of an antiproton and an antineutron) are believed to be quite promising as a tool for DM searches. 
They can be produced by DM via the coalescence of an antiproton and an antineutron originating from an annihilation event, provided that the latter ones are produced with momenta that are spatially aligned and comparable in magnitude~\cite{Donato:1999gy}. This peculiar kinematics in the production mechanism implies two things. One, that the flux of $\bar d$ from DM is (unfortunately) predicted to be much lower than the one of other, more readily produced, charged CRs. Two, that (fortunately) the flux peaks in an energy region, corresponding typically to a fraction of a GeV, where very little astrophysical background $\bar d$'s are present, since the latter ones are believed to originate in spallations of high energy cosmic ray protons on the interstellar gas at rest, a completely different kinematical situation. It is therefore sometimes said that the detection of even just one sub-GeV antideuteron in CRs would be a smoking gun evidence for DM. 

At ICRC 2015, antideuterons have been discussed within the context of the {\sc Gaps} and {\sc Ams-02} experiments~\cite{vonDoetinchem,vonDoetinchem:2015zva}. The former one, in particular, will be a dedicated balloon mission which employs a novel technique: it plans to slow down the $\bar d$ nucleus, have it captured inside the detector to form an exotic atom and then annihilate emitting characteristic X-ray and pion radiation. Currently there is only an upper limit from the {\sc Bess} experiment~\cite{BESSlimit}, at the level of about 2 orders of magnitude higher than the most optimistic predictions. The limit and the reach of the future experiments is reported in fig.~\ref{fig:antideuteron}.

\section{Indirect detection via gamma rays}
\label{sec:gammas}


\noindent In general, DM annihilation (or decay) can produce photon fluxes in many ways, among which:
\begin{itemize}
\item[I)] `Prompt' gamma-rays: produced directly by DM annihilations themselves. In turn, however, these gamma-rays can originate from different stages of the annihilation process:
\begin{itemize}
\item[Ia)] From the bremsstrahlung of charged particles and the fragmentation of hadrons, e.g. $\pi^0$, in the final states of the annihilations. These processes generically give origin to a {\it continuum} of $\gamma$-rays which peaks at energies somewhat smaller than the DM mass $m_{\mbox{\tiny DM}}$, i.e.\ typically in the range of tens of GeV to multi-TeV. The spectra can be computed in a model independent way (see e.g.~\cite{Cirelli:2010xx}), since all one needs to know is the pair of primary SM particles. \label{promptcontinuum}
\item[Ib)] From the {\it bremsstrahlung} from one of the {\it internal} particles in the annihilation diagram. This typically gives rise to a sharp feature that peaks at an energy corresponding to the DM mass. The process is in general subdominant with respect to the continuum, but it can be particularly important in cases in which the continuum itself is suppressed by some mechanism, e.g. helicity constraints, which are lifted by the internal radiation. The spectrum from this contribution cannot be computed without knowing the details of the annihilation model. \label{promptIB}
\item[Ic)] From an annihilation directly into a pair of gamma-rays, which gives rise to a {\it line} spectrum at the energy corresponding to the mass of the DM. Since DM is neutral, this annihilation has to proceed via some intermediation (typically a loop of charged particles) and it is therefore suppressed by (typically) 2 to 4 orders of magnitude. \label{promptline}
\end{itemize}
In any case, since these $\gamma$-rays originate directly from the annihilations themselves, their spatial distribution follows closely the distribution of DM.

\item[II)] Secondary radiation, emitted by the $e^+e^-$, which have been produced by the annihilation process, when they interact with the environment:
\begin{itemize} 
\item[IIa)] ICS gamma-rays: produced by the Inverse Compton Scattering (ICS) of the energetic electrons and positrons, created in the DM annihilation, onto the low energy photons of the CMB, the galactic star-light and the infrared-light, which are thus upscattered in energy. Typically, they cover a wider range of energies than prompt gamma rays, from energies of a fraction of the DM mass to almost up to the DM mass itself. Their spatial distribution traces the distribution of $e^\pm$, which originate from DM but then diffuse out in the whole galactic halo (as seen above).

\item[IIc)] Bremsstrahlung gamma-rays: produced by the same energetic electrons and positrons onto the gas in the interstellar medium. Typically, they are of lower energy than ICS rays; their spatial distribution traces the $e^\pm$ but also the density of target gas, so it is typically concentrated in the galactic disk.

\item[IIc)] Synchrotron emission: consisting in the radiation emitted in the magnetic field of the Galaxy by the $e^\pm$ produced by DM annihilations. For an intensity of the magnetic field of $\mathcal{O}$($\mu$Gauss), like in the case of the Milky Way halo, and for $e^\pm$ of GeV-TeV energies, the synchrotron emission falls in the MHz-GHz range, i.e. in the radio band. For large magnetic fields and large DM masses it can reach up to EHz, i.e. the X-ray band. Their region of origin is necessarily concentrated where the magnetic field is highest; in particular the GC is the usual target of choice. However, it has been recently suggested that the galactic halo at large, or even the extragalactic ones, can be interesting sources~\cite{Fornengo:2011iq}.
\end{itemize}
\end{itemize}

\noindent Individuating the best targets to search for these annihilation signals is one of the main games in the field. Not very surprisingly, the preferred targets have to be {\bf (i)} regions with high DM densities {\em and/or}\  {\bf (ii)} regions where the astrophysical `background' is reduced and therefore the signal/noise ratio is favorable. The distinction between (i) and (ii) is of course not clear-cut, and of course there are specific cases in which other environmental reasons make a region more suitable than another (such as in the case of synchrotron radiation which needs a region with a strong magnetic field). Moreover, new promising targets keep being individuated. However, for the sake of schematizing, one can list the following targets at which the experiments look:
\begin{itemize}
\item[$\circ$] The Milky Way Galactic Center (GC) $-$ {\bf (i)}. 
\item[$\circ$] Small regions around or just outside the GC, such as the Galactic Center Halo (GCH, an annulus of about $1^\circ$ around the GC, excluding the Galactic Plane) etc $-$ {\bf (i)} + {\bf (ii)}.
\item[$\circ$] Wide regions of the Galactic Halo (GH) itself (such as observational windows centered at the GC and several tens of degrees wide in latitude and longitude, from which a diffuse flux of gamma-rays is expected, including the one due to the ICS emission from the diffused population of $e^\pm$ from DM annihilations $-$ {\bf (ii)}.
\item[$\circ$] Globular clusters (GloC), which are dense agglomerates of stars, embedded in the Milky Way galactic halo. They are a peculiar kind of target since they are not supposed to be DM dominated, quite the opposite, as they are rich of stars. The interest in them arises from two facts: that they may have formed inside a primordial DM subhalo and some of the DM may have remained trapped; that the density of baryonic matter may create by attraction a DM spike and thus enhance the annihilation flux$-$ {\bf (i)}. 
\item[$\circ$] Subhalos of the galactic DM halo, whose position, however, is of course unknown a priori. 
\item[$\circ$] Satellite galaxies of the Milky Way, often of the dwarf spheroidal (dSph) class, such as Sagittarius, Segue1, Draco and several others, which are star-deprived and believed to be DM dominated $-$ {\bf (i)} + {\bf (ii)}. The `problem' with dSphs is precisely that the determination of their DM content and distribution relies on stars as kinematical tracers and therefore suffers from rather large uncertainties~\cite{bonnivard}.
\item[$\circ$] Large scale structures in the relatively nearby Universe, such as galaxy clusters (e.g. the Virgo, Coma, Fornax, Perseus clusters, and several others with catalog names that are less pleasant to write) $-$ {\bf (i)} + {\bf (ii)}
\item[$\circ$] The Universe at large, meaning looking at  the {\em isotropic} flux of (redshifted) $\gamma$-rays that come to us from DM annihilation in all halos and all along the recent history of the Universe. Often this flux is called `extragalactic' or `cosmological' $-$ {\bf (ii)}
\end{itemize}

\begin{figure}[p]
\small
\begin{center}

\begin{tabular}{|l||c|c|c|c|c|c|}
\multicolumn{7}{l}{\bf Gamma ray searches - DM annihilation } \\
\hline
 & {\bf GC / GCH} & {\bf MW halo} & {\bf Dwarfs} & {\bf Clusters} & {\bf Extragalactic} & {\bf other} \\
\hline \hline
{\multirow{6}{*}{\rotatebox[origin=c]{90}{\bf continuum }}} & {\sc Hess}~\cite{Abramowski:2011hc,Lefranc:2015vza} & {\sc Fermi}~\cite{Ackermann:2012rg} & {\sc Magic}~\cite{Aleksic:2013xea} & {\sc Fermi}~\cite {Ackermann:2010rg,Ackermann:2015fdi} & {\sc Fermi}~\cite{Ackermann:2015tah} & {\sc Fermi}~\cite{Ackermann:2012nb} \\
 & ({\sc Fermi}~\cite{TheFermi-LAT:2015kwa}) & & {\sc Hess}~\cite{Abramowski:2014tra} & {\sc Hess}~\cite{Abramowski:2012au} & & {\footnotesize (dark satellites)} \\\cline{7-7}
 & & & {\sc Fermi}~\cite{Ackermann:2015zua} & {\sc Veritas}~\cite{Pfrommer:2012mm} & &  {\sc Hess}~\cite{Abramowski:2011hh} \\ 
 & & & {\sc Fermi}+{\sc Magic}~\cite{Rico:2015nya} & & & {\footnotesize (GloCs)} \\ \cline{7-7}
 & & & {\sc Fermi-Des}~\cite{Drlica-Wagner:2015xua} & & &  {\sc Veritas}~\cite{NietoCastano}  \\ 
 & & & {\sc Hawc}~\cite{Harding:2015bua} & & & {\footnotesize (subhalos) }\\  
 & & & {\sc Veritas}~\cite{Zitzer:2015eqa} & & & \\
\hline
\rotatebox[origin=c]{90}{\bf  \ lines } & {\sc Hess}~\cite{Abramowski:2013ax,Kieffer:2015nsa} & {\sc Fermi}~\cite{Ackermann:2015lka} & {\sc Magic}~\cite{Aleksic:2013xea} & {\sc Fermi}~\cite{FERMIclusterslines}& & \\
\hline
\end{tabular}
\caption{\label{tab:gammaconstraints} A collection of the currently most relevant $\gamma$-ray searches for DM annihilation, as produced by the different experimental collaborations. }

\medskip

\includegraphics[width= 0.47 \columnwidth]{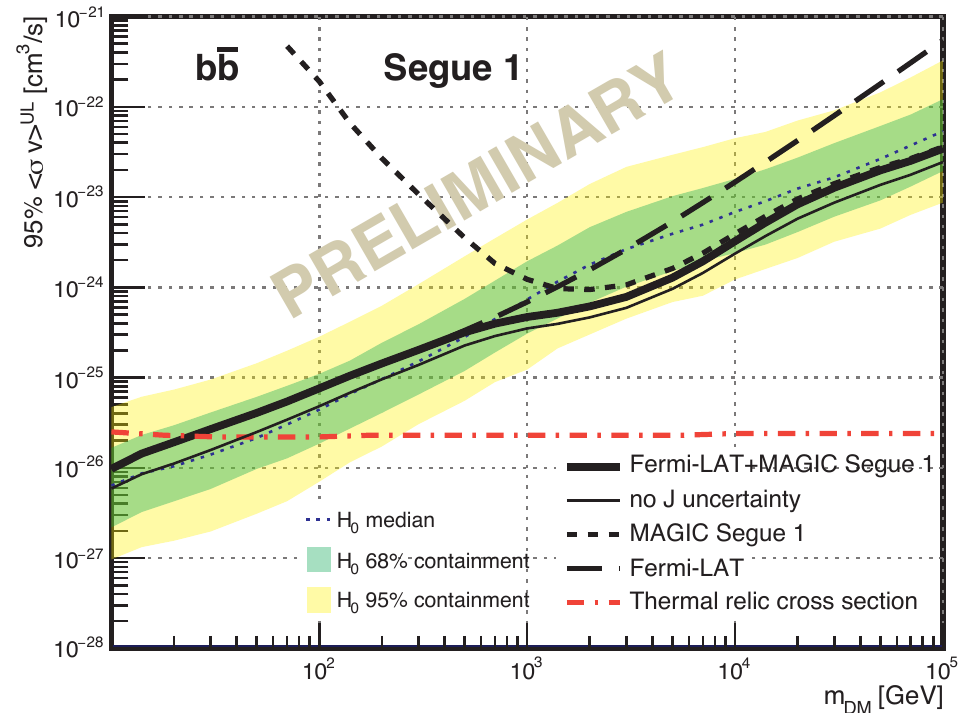} 
\includegraphics[width= 0.52 \columnwidth]{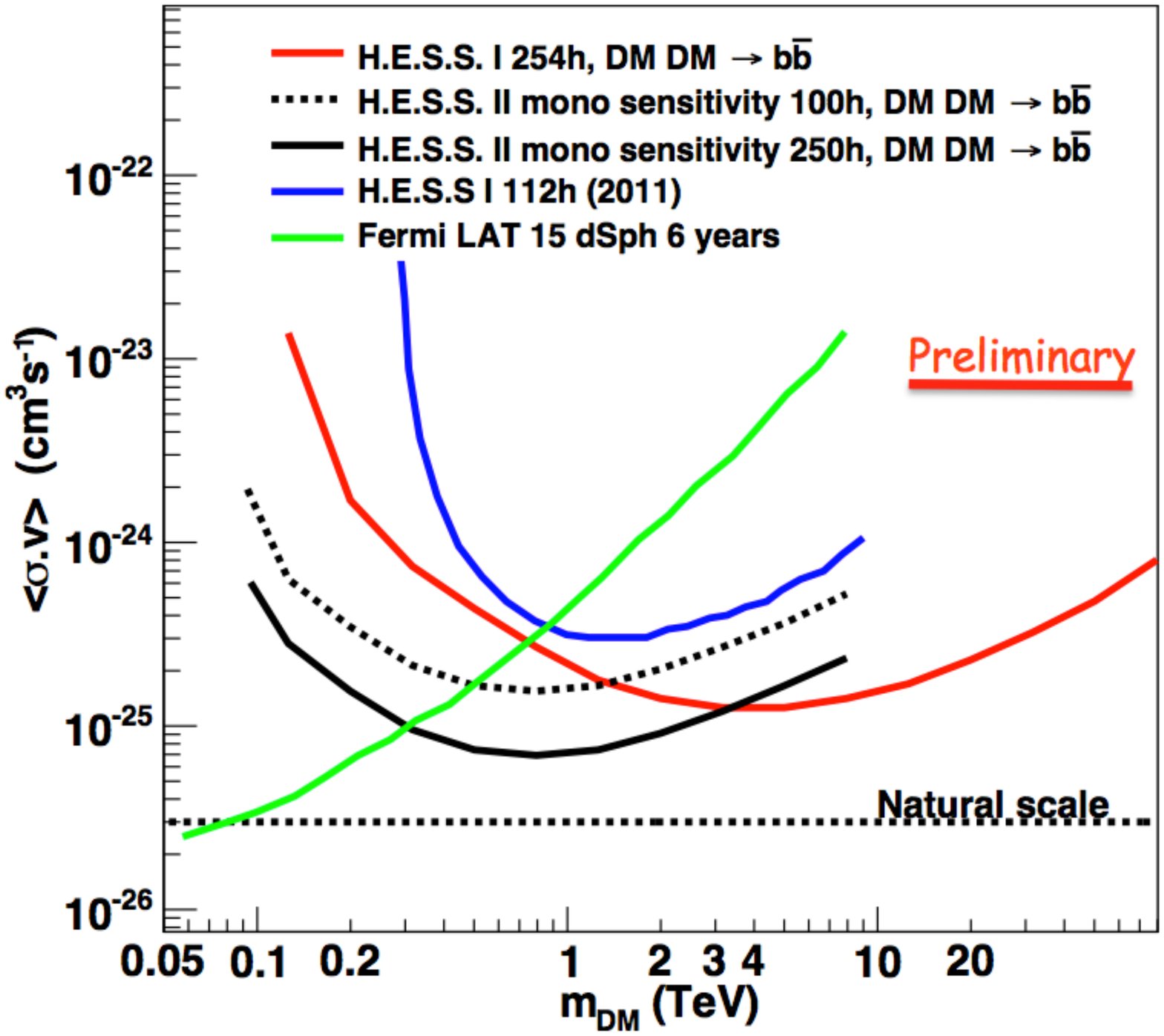} 
\caption{\label{fig:othergamma} Two new results concerning $\gamma$-ray constraints on DM that have been presented at ICRC 2015: the bound by {\sc Fermi} and {\sc Magic} from the observation of dwarf galaxies (left, from~\cite{Rico:2015nya}) and the bounds by {\sc Hess} from the observation of the GC region (right, from~\cite{Lefranc:2015vza}).}

\bigskip

\includegraphics[width= 0.78 \columnwidth]{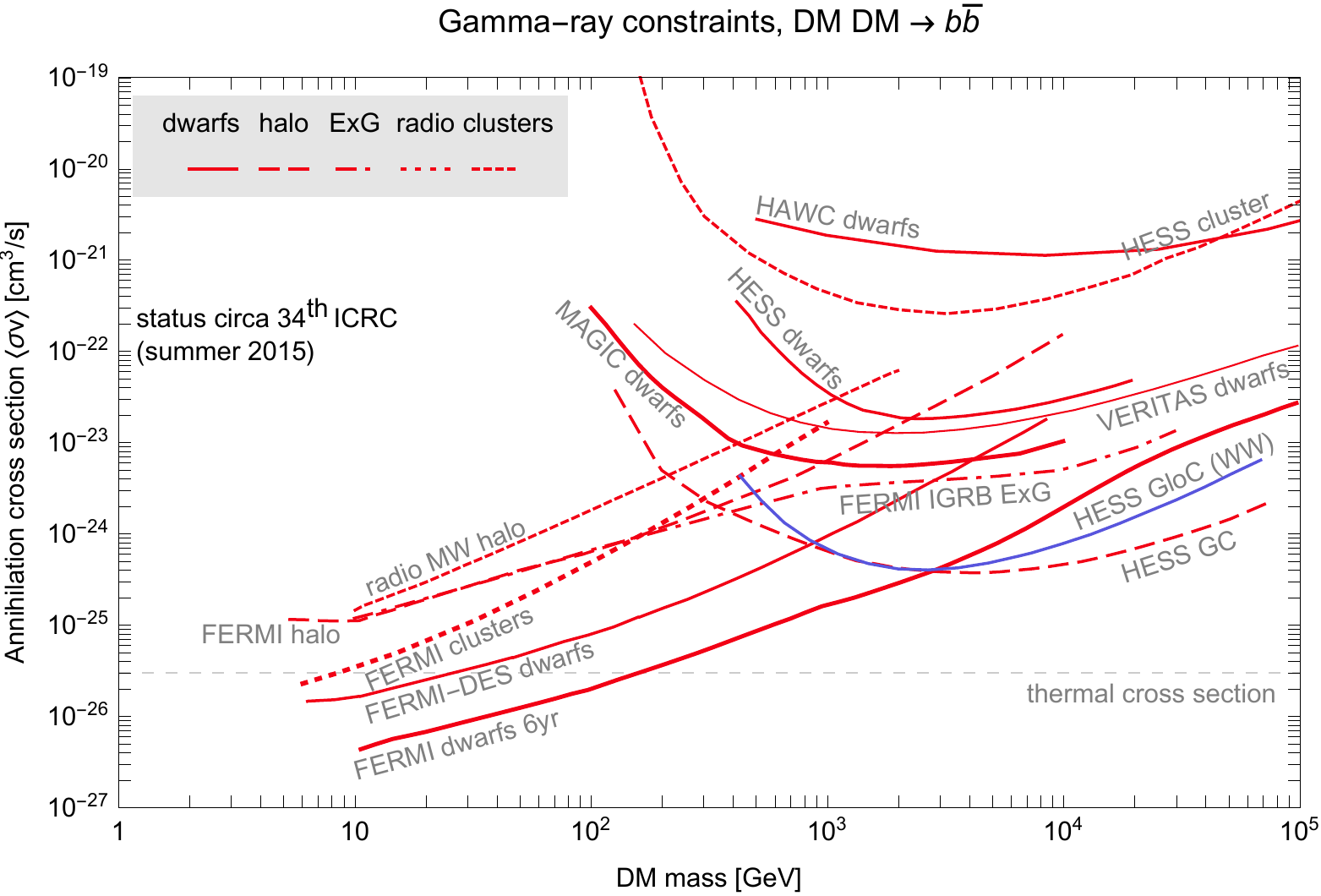} 
\caption{Bounds on DM annihilation imposed by different gamma-ray (and radio) observations.}
\label{fig:all_ID_constraints_gamma}
\end{center}
\end{figure}

\noindent Focussing on the range of energies above a GeV or so (i.e. proper gamma rays), the current main experiments in the game are the {\sc Fermi} satellite and the ground-based Imaging Atmospheric \v Cerenkov Telescopes (IACT) {\sc Hess}, {\sc Magic} and  {\sc Veritas}. Table~\ref{tab:gammaconstraints} lists the most relevant searches for signals from DM annihilation performed by these experiments: they all turn out empty handed and therefore bounds on the annihilation cross section are imposed. 
Moreover, equivalent searches are performed for DM decay. As an example, at ICRC 2015 {\sc Magic} presented a limit from the observation of the Perseus cluster~\cite{Palacio:2015nza}.
Two bounds presented at ICRC 2015 for the first time are reported in fig.~\ref{fig:othergamma} (they actually turn out to be the most stringent ones across the whole range of masses). Fig.~\ref{fig:all_ID_constraints_gamma} presents instead the whole collection of limits (choosing the continuum $\gamma$-rays from the $b \bar b$ annihilation channel for definiteness, although the GloC bounds of {\sc Hess} apply to the $W^+W^-$ case), mostly corresponding to those listed in table of fig.~\ref{tab:gammaconstraints}. 
The radio bounds are taken from~\cite{Fornengo:2011iq}.

Besides these `official' searches, many independent works have analysed varying combinations of datasets, considered different targets and studied different models, often obtaining more stringent limits. For the purposes of this write-up, I choose not to enter this very large body of literature, which is however extremely relevant and interesting.
 

\medskip

At ICRC 2015, plans for the future of gamma-ray detection in the DM context have been discussed too. The main players are expected to be the \v Cerenkov Telescope Array ({\sc Cta})~\cite{Carr}, the High Altitude Water \v Cerenkov Observatory ({\sc Hawc})~\cite{Harding} and the Large High Altitude Air Shower Observatory ({\sc Lhaaso})~\cite{DiSciascio}.

\subsection{The Galactic Center GeV excess}
\label{sec:GeVexcess}

\begin{figure}[t]
\includegraphics[width= 0.56 \columnwidth]{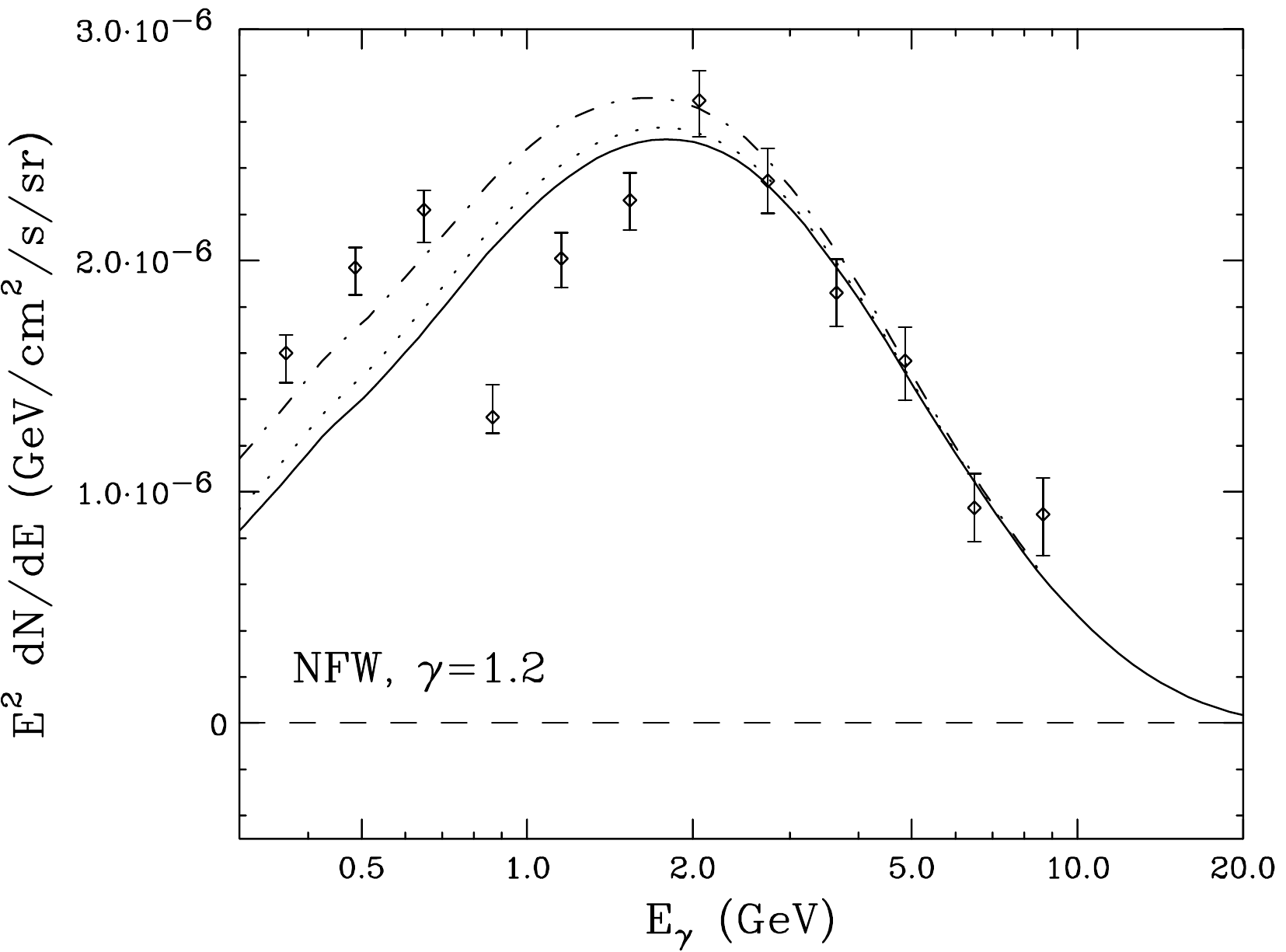} \quad
\includegraphics[width= 0.42 \columnwidth]{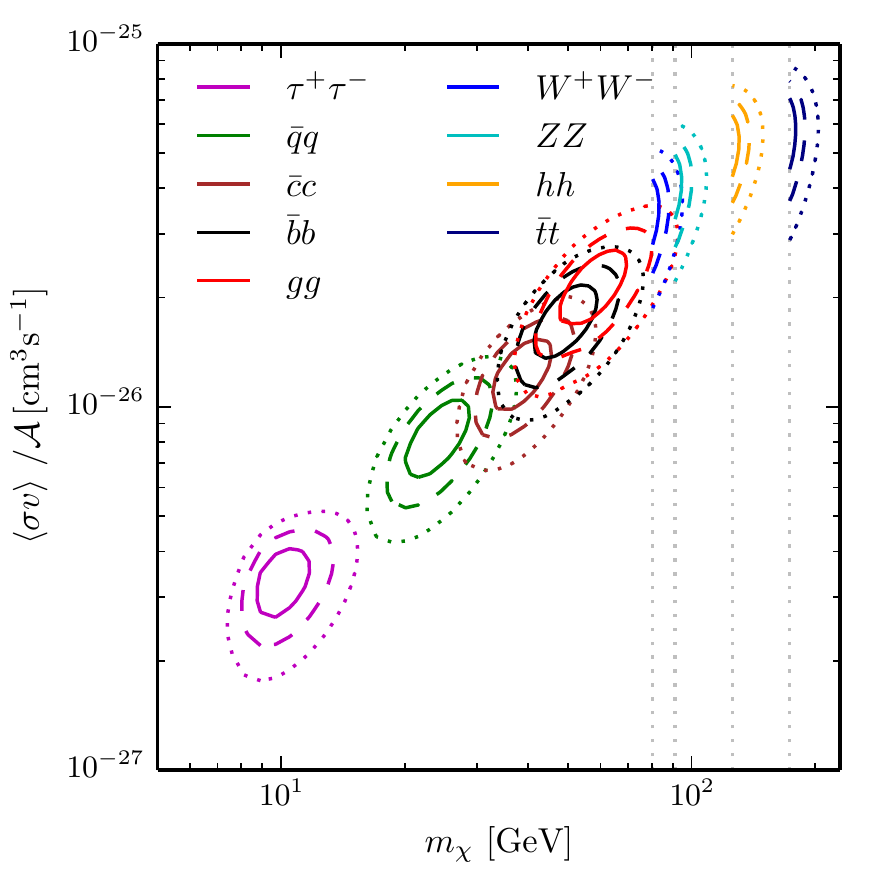} 
\caption{\label{fig:GeVexcess} The GC GeV excess as seen in residual {\sc Fermi} data (left, in the analysis of~\cite{Daylan:2014rsa}) and its DM interpretation (right, from~\cite{Calore:2014nla}).}

\bigskip

\includegraphics[width= 0.33 \columnwidth]{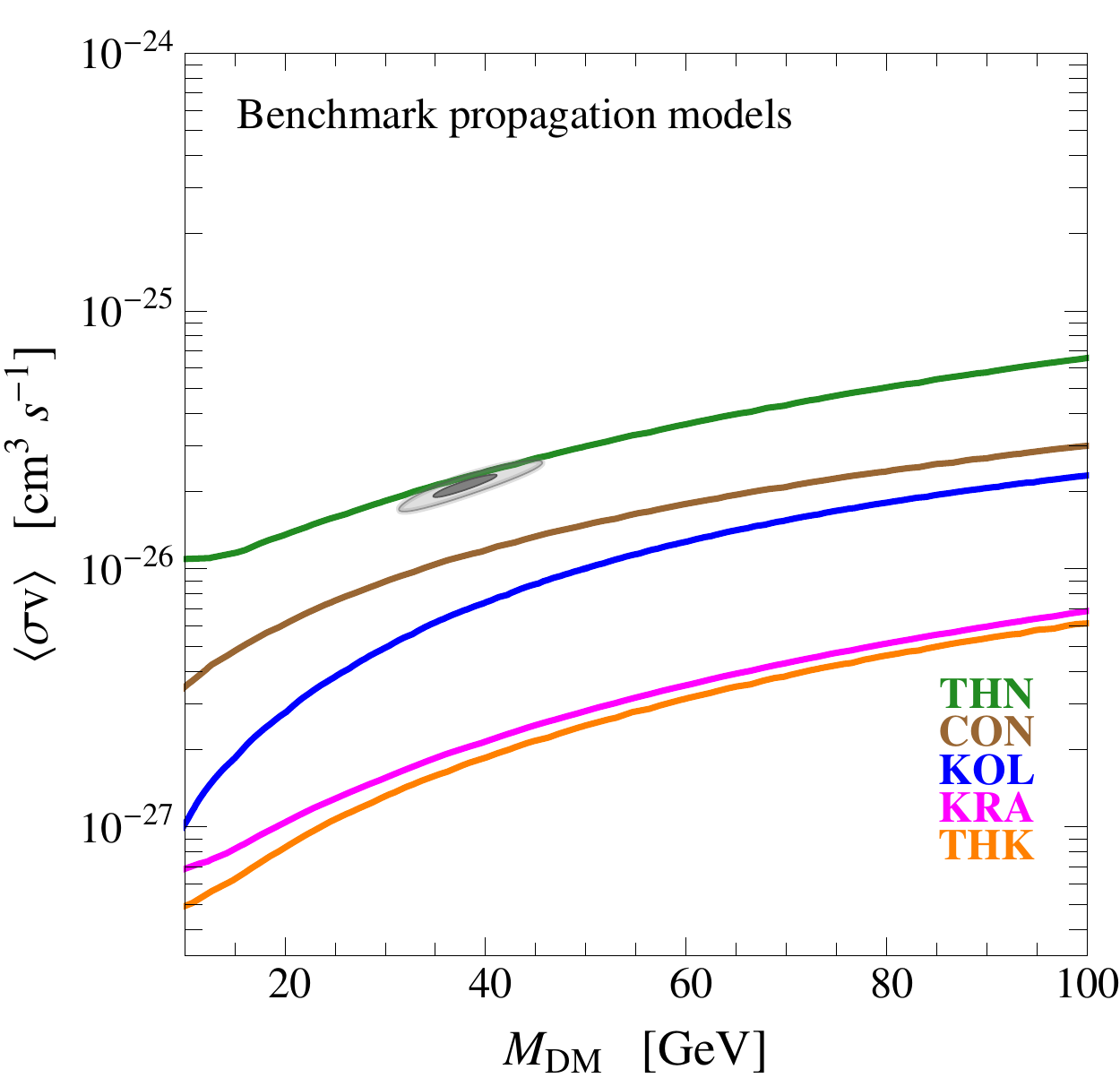} 
\includegraphics[width= 0.33 \columnwidth]{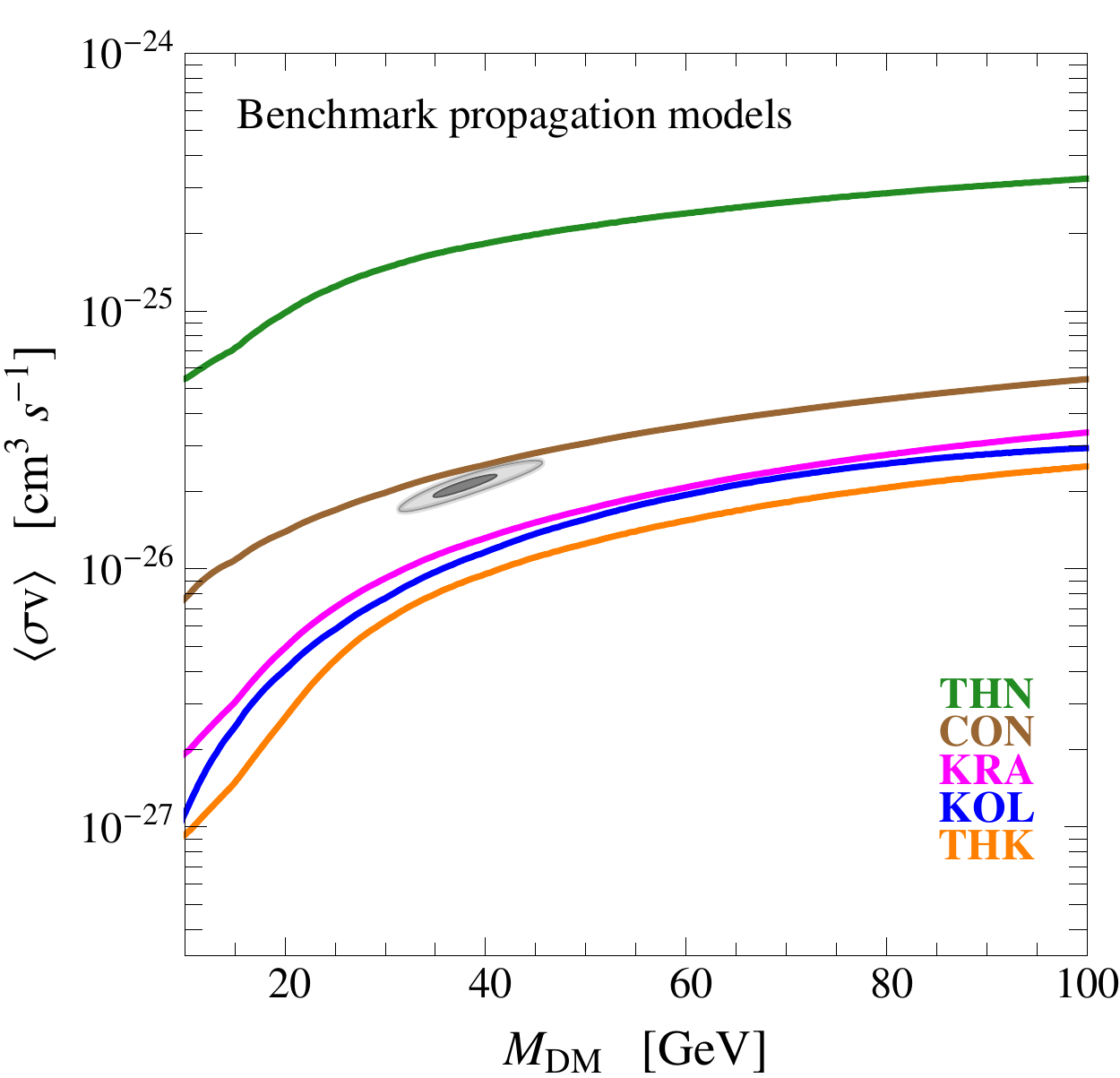} 
\includegraphics[width= 0.33 \columnwidth]{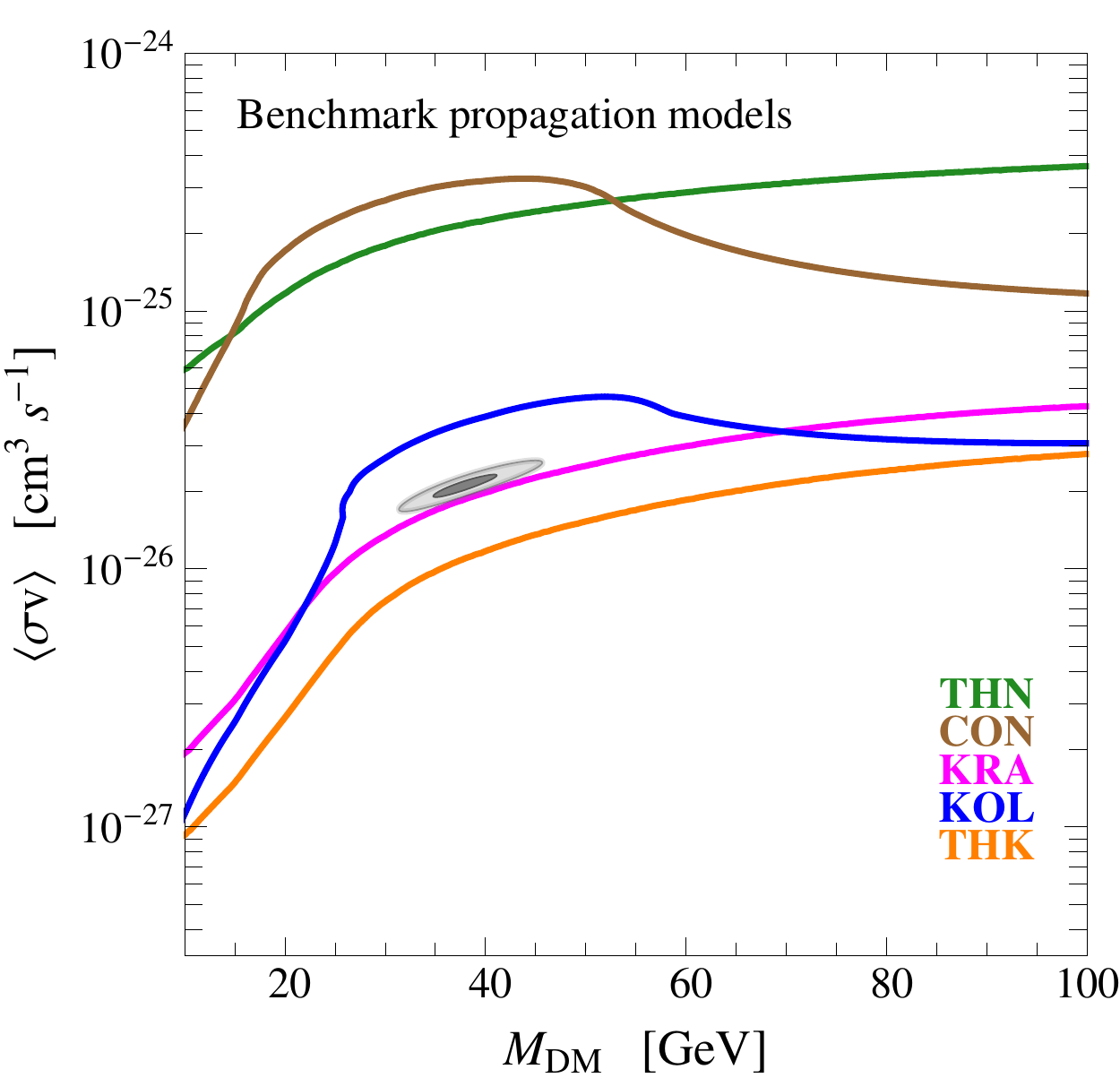} 
\caption{\label{fig:GeVexcessconstraints} Antiproton constraints on the DM interpretation of the GC GeV excess, assuming fixed (left), flexible (middle) or essentially free (right) parameters for solar modulation (from~\cite{Cirelli:2014lwa} or \cite{Gaggero:2015mga}). }

\end{figure} 

Possibly the most hotly debated issue concerning DM at ICRC 2015 is the Galactic Center GeV excess. The issue has relatively old origins: several authors have reported, since 2009, the detection of a gamma-ray signal in {\sc Fermi} data, originating from the inner few degrees around the GC~\cite{hooperon_history}. The spectrum and the morphology are compatible with those expected from annihilating DM particles, and in particular they are best fit by 30-50 GeV DM particles distributed according to a (contracted) NFW profile and annihilating into $b\bar{b}$ with $\langle \sigma v\rangle = 1.4 \div 2 \times 10^{-26}$ cm$^3/{\rm s}$~\cite{Daylan:2014rsa} (see fig.~\ref{fig:GeVexcess}). This has naturally given rise to an intense model building actvity~\cite{hooperon model building} (including on the possibility of tests at colliders~\cite{Buchmueller:2015eea}).
A `semi-official' statement from the {\sc Fermi} collaboration on this issue has been given at ICRC 2015~\cite{Porter:2015uaa} and then in~\cite{TheFermi-LAT:2015kwa}: a residual is indeed present and it can be well fit by DM as discussed above, although any conclusion is premature. 
The presence of the excess (and its DM interpretation) is robust against variations of the {\em known} astrophysical foregrounds~\cite{astrovariations,Calore:2014nla}. But of course, as usual, the excess might be due to {\em unknown} (up to now) astrophysical phenomena: millisecond pulsars or other point sources~\cite{MSP,Bartels:2015aea,Weniger,Lee:2015fea,O'Leary:2015gfa}, leptonic or other outbursts~\cite{Cholis:2015dea,Calore,Petrovic:2014uda,Carlson:2014cwa}, non-conventional CR propagation~\cite{1411.7623} or injections of CRs~\cite{Gaggero:2015nsa,Urbano,Carlson:2015ona}.

A relevant question is whether the DM interpretation can be confirmed or ruled out by associated signals, for instance in antiprotons (as asked in~\cite{Cirelli:2014lwa,Gaggero,Bringmann:2014lpa,Hooper:2014ysa,Ipek:2014gua}) or in gamma rays from dwarfs~\cite{wood,Ackermann:2015zua,Abazajian:2015raa}. 
In both cases the conclusions seem to vary depending on the details of the analysis which is done (see fig.~\ref{fig:GeVexcessconstraints}). Hence, the conservative bottom line is still that no, the constraints are not conclusive in either sense.

\section{Indirect detection via neutrinos}
\label{sec:neutrinos}

\begin{figure}[!p]
\begin{center}
\includegraphics[width= 0.43 \columnwidth]{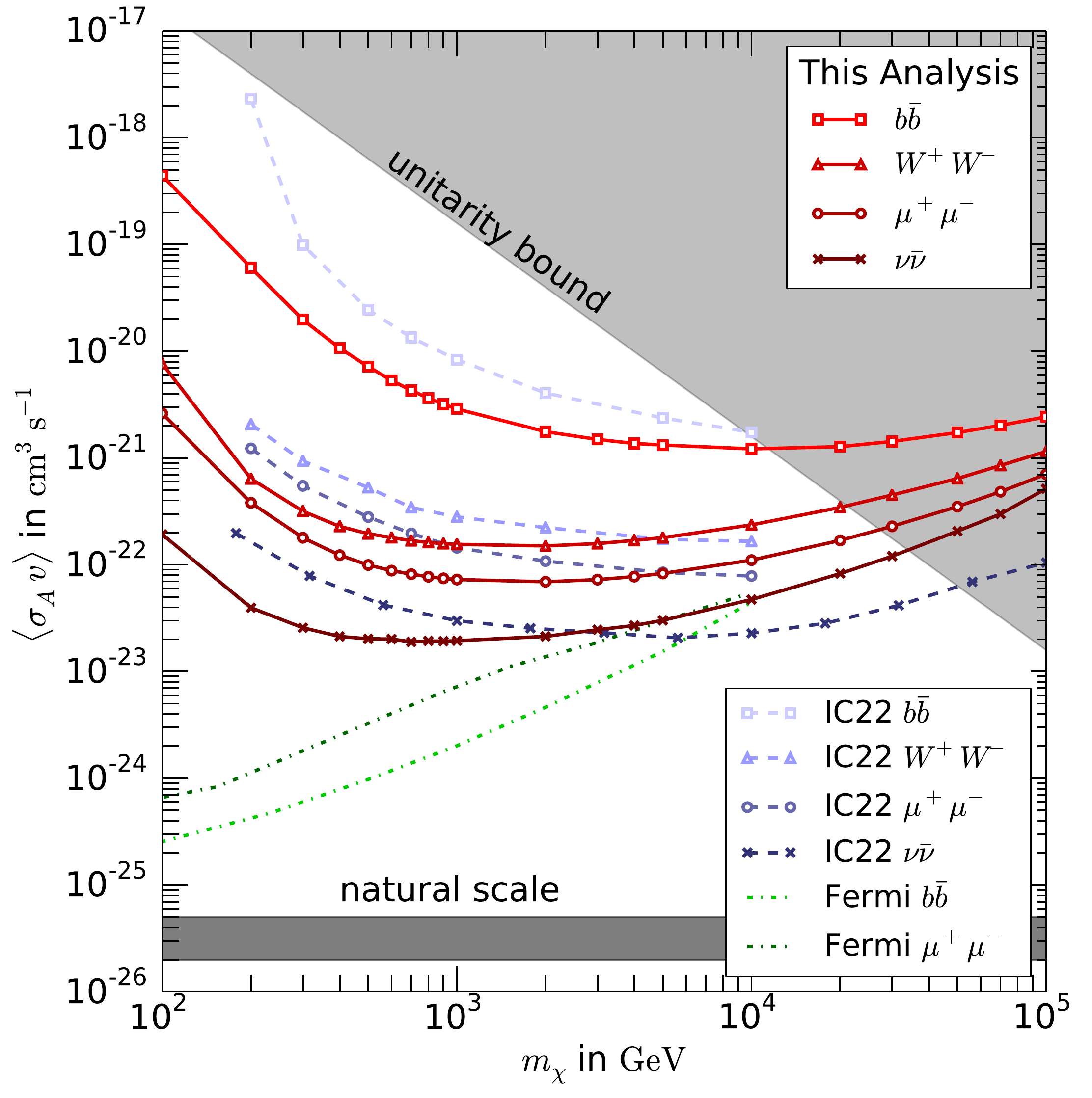} 
\includegraphics[width= 0.56 \columnwidth]{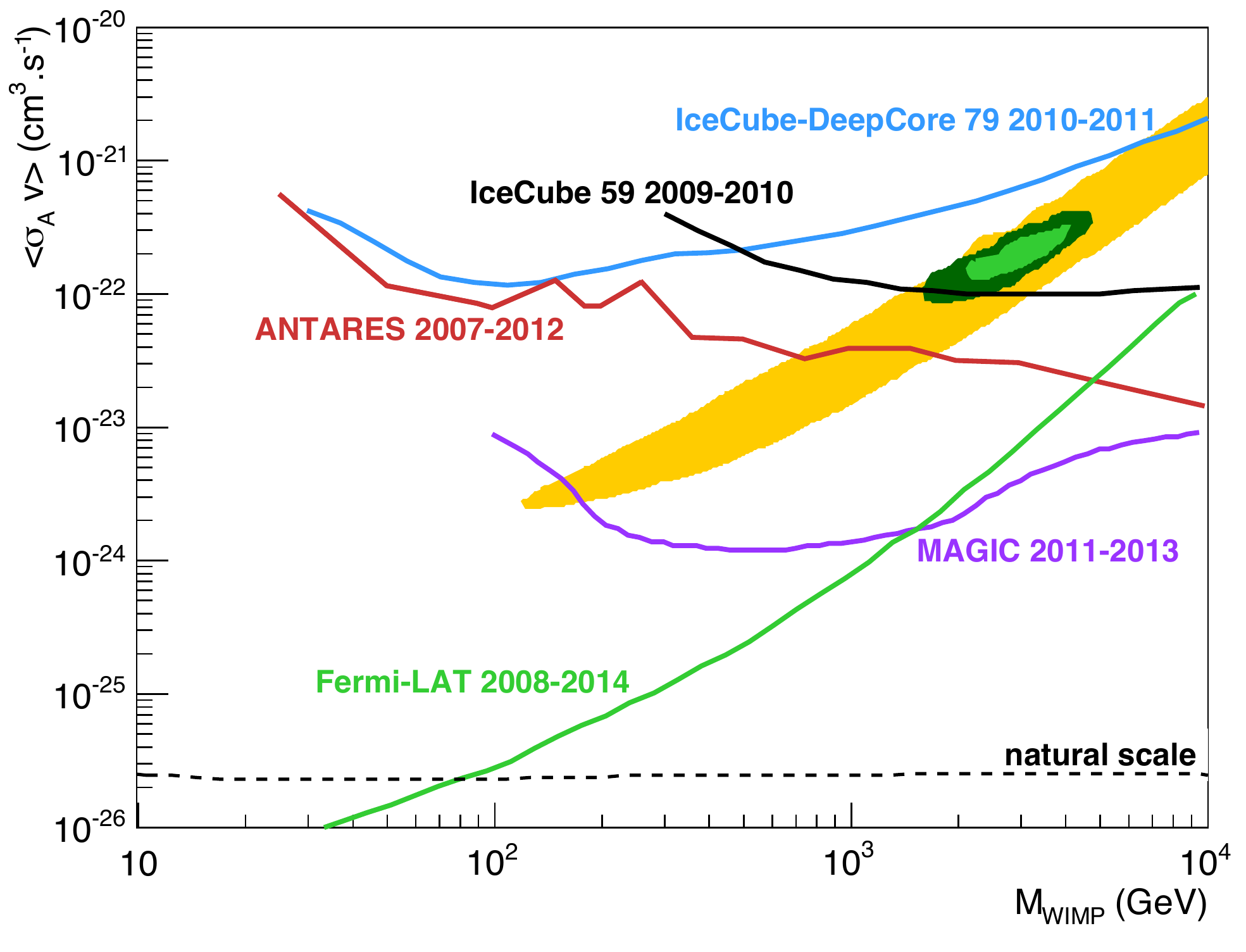} \\[3mm]
\includegraphics[width= 0.49 \columnwidth]{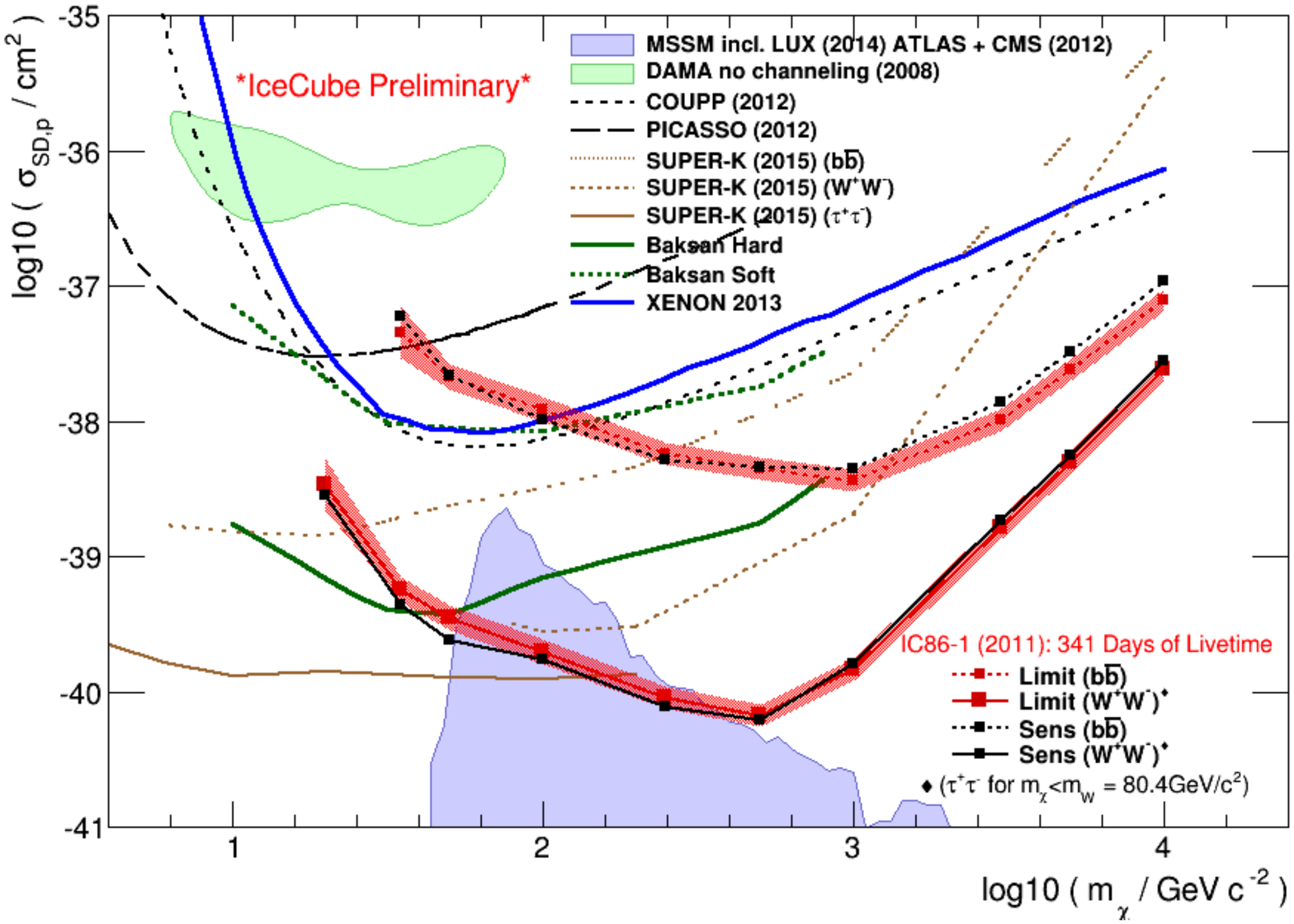} 
\includegraphics[width= 0.49 \columnwidth]{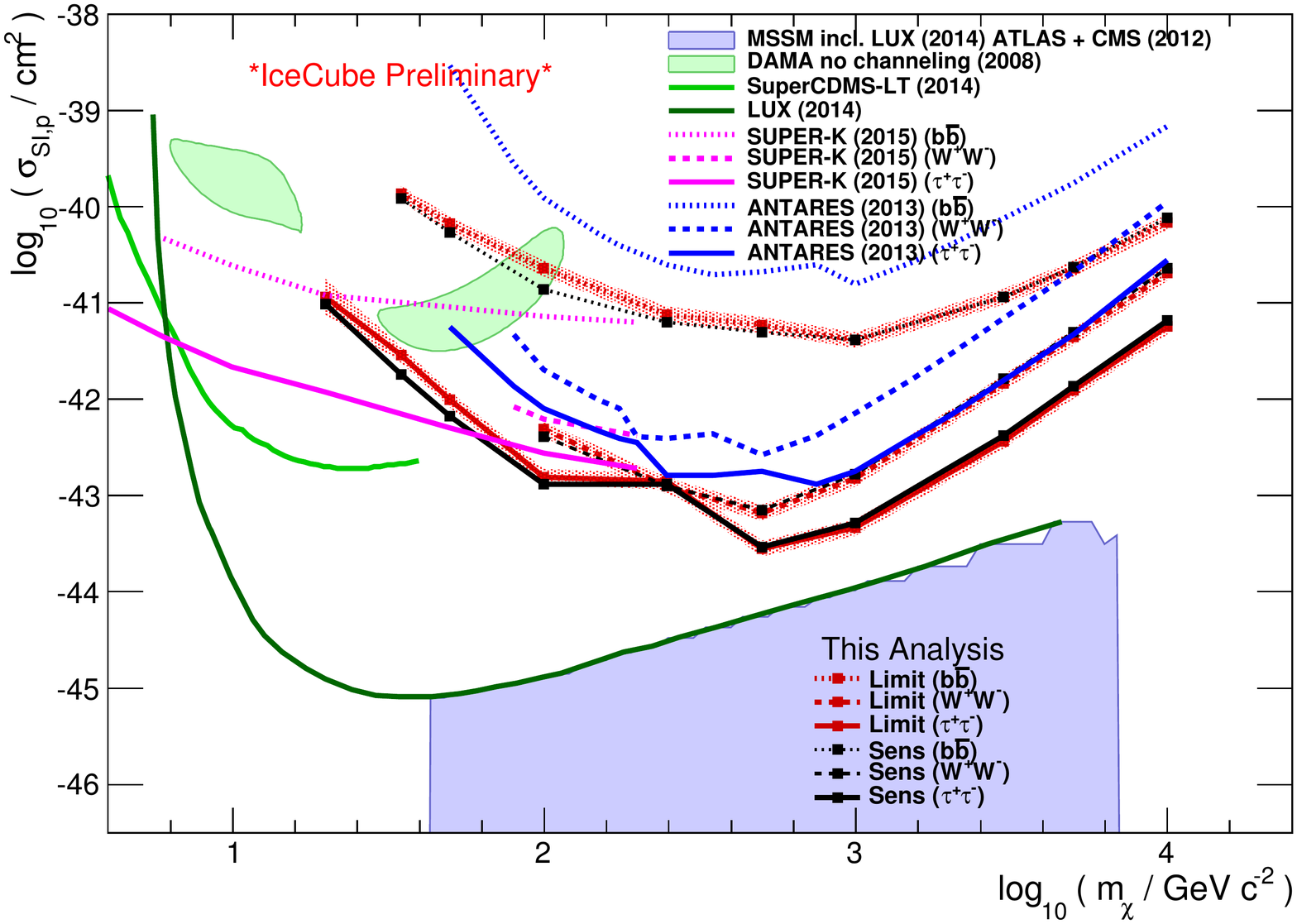} \\[3mm]
\includegraphics[width= 0.95 \columnwidth]{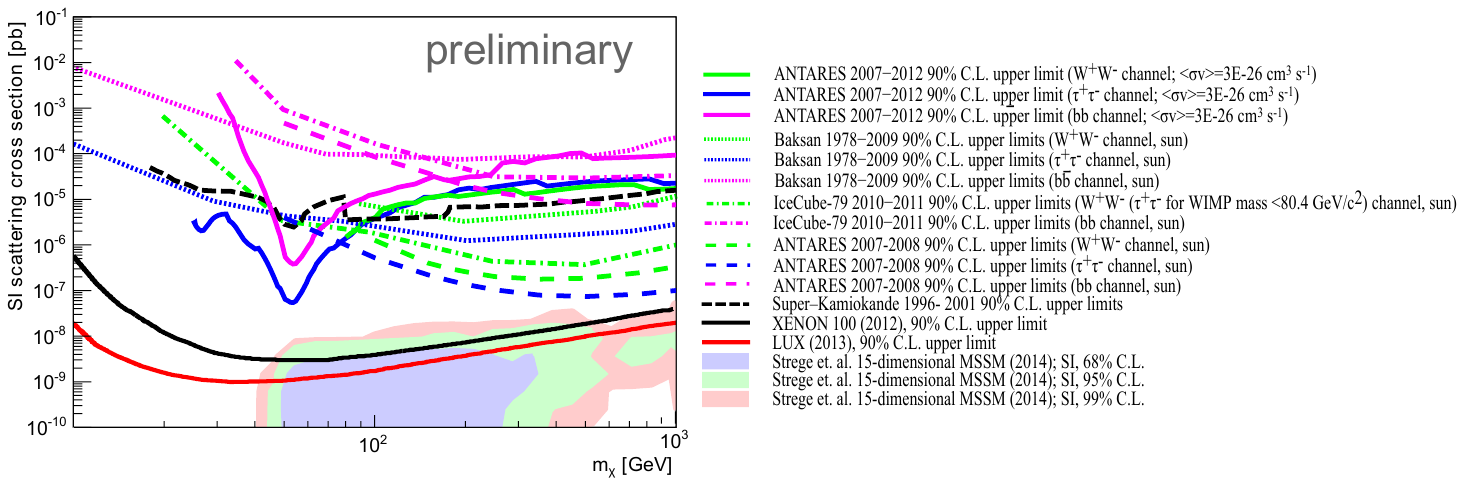}
\caption{A collection of current neutrino bounds on Dark Matter. First line: bounds on DM annihilations in the Galactic Halo (left, figure from {\sc Icecube}~\cite{Aartsen:2014hva}) or the Galactic Center (right, figure from {\sc Antares}~\cite{Adrian-Martinez:2015wey}, referring to the $\tau^+\tau^-$ channel). Notice that the left figure assumes a NFW profile renormalized with a local density of 0.47 GeV/cm$^3$ while the right one assumes 0.4. Second line: bounds on the DM scattering cross section with nucleons, from DM accumulated in the center of the Sun, obtained by {\sc Icecube} (left: figure from~\cite{Rameez}, for the Spin Dependent case; right: figure from the parallel analysis in~\cite{Zoll}, for the Spin Independent case). Third line: bounds on the SI cross section from DM in the center of the Earth (figure from~\cite{Gleixner}).}
\label{fig:neutrinos}
\end{center}
\end{figure}

Neutrinos are of course produced in DM annihilations together with all the other particles discussed above. Similarly to $\gamma$-rays, neutrinos have the advantage of proceeding straight and essentially unabsorbed through the Galaxy. Even more, they can cross long lengths of dense matter with little interaction. Contrary to $\gamma$-rays, however, the detection principle of neutrinos is more difficult and it introduces limitations in the choice of targets. Neutrinos are observed at huge \v Cerenkov detectors located underground (or under-ice or under-water) via the showers of secondary particles that they produce when interacting in the material inside the instrumented volume or in its immediate surroundings. The charged particles, in particular muons, emit \v Cerenkov light when traversing the experiment and thus their energy and direction (which are connected to those of the parent neutrino) can be measured. The main background for this search consists in the large flux of cosmic muons coming from the atmosphere above the detector. The experiments, therefore, have to select only upgoing tracks, i.e. due to neutrinos that have crossed the entire Earth and interacted inside or just below the instrumented volume. 
Schematically, experiments look for neutrinos:
\begin{itemize}
\item[$\circ$] From the GC or the GH, in close similarity with $\gamma$-rays. Experiments located at the South Pole can not `see' the GC, which is essentially above horizon for them. The DeepCore extension of {\sc Icecube}, however, circumvents this limitation by using the outer portion of the experiment as an active veto.
\item[$\circ$] From satellite galaxies or clusters of galaxies, again in similarity with $\gamma$-rays, although in this case the sensitivities are typically not competitive with gamma rays.
\item[$\circ$] From the center of the Sun (or even the Earth). The idea is that DM particles in the halo may become gravitationally captured by a massive body, lose energy via repeated scatterings with its nuclei and thus accumulate at its center. The annihilations occurring there give origin to fluxes of high energy neutrinos which, albeit suffering oscillations and interactions in the dense matter of the astrophysical body (see e.g.~\cite{DMnu,Baratella:2013fya,Blennow:2007tw}), can emerge. The detection of high-energy neutrinos from the Sun, on top of the much lower energy neutrino flux due to nuclear fusion processes, would constitute the proverbial smoking gun for DM, as there are no known astrophysical processes able to mimic it.
\end{itemize}

\medskip

At ICRC 2015, neutrinos from DM have been discussed in a number of contributions~\cite{Rameez,Zoll,Gleixner,Kunnen,Ardid,dewith}. 
The main neutrino telescopes, such as {\sc SuperKamiokande}, {\sc Icecube} and {\sc Antares}, have looked for signals, without finding any. This therefore imposes once again bounds on the relevant DM properties. Fig.~\ref{fig:neutrinos} collects a few representative ones. 
The non-observation of high-energy neutrino fluxes from the Galactic Halo and Galactic Center imposes constraints on the DM annihilation cross section (first line of fig.~\ref{fig:neutrinos}). Quantitatively, such constraints fall somewhat above the $\gamma$-ray bounds discussed in Sec.~\ref{sec:gammas}, and are therefore slightly less stringent. They have the advantage, however, of being less dependent on the DM particle mass: the reason is that while a larger DM mass implies lower DM density (e.g. in the GC) and therefore fainter fluxes and looser constraints, this is compensated by the fact that the higher energy neutrinos coming from such heavier DM annihilations also have a higher cross section for detection (in the material of neutrino telescopes) and thus the loss in the rate is partly compensated. Neutrino constraints become therefore somewhat competitive with $\gamma$-ray bounds at large DM masses. Constraints from the observations of dwarf galaxies are for the moment less competitive~\cite{dewith}.

\begin{figure}[t]
\begin{center}
\includegraphics[width= 0.98 \columnwidth]{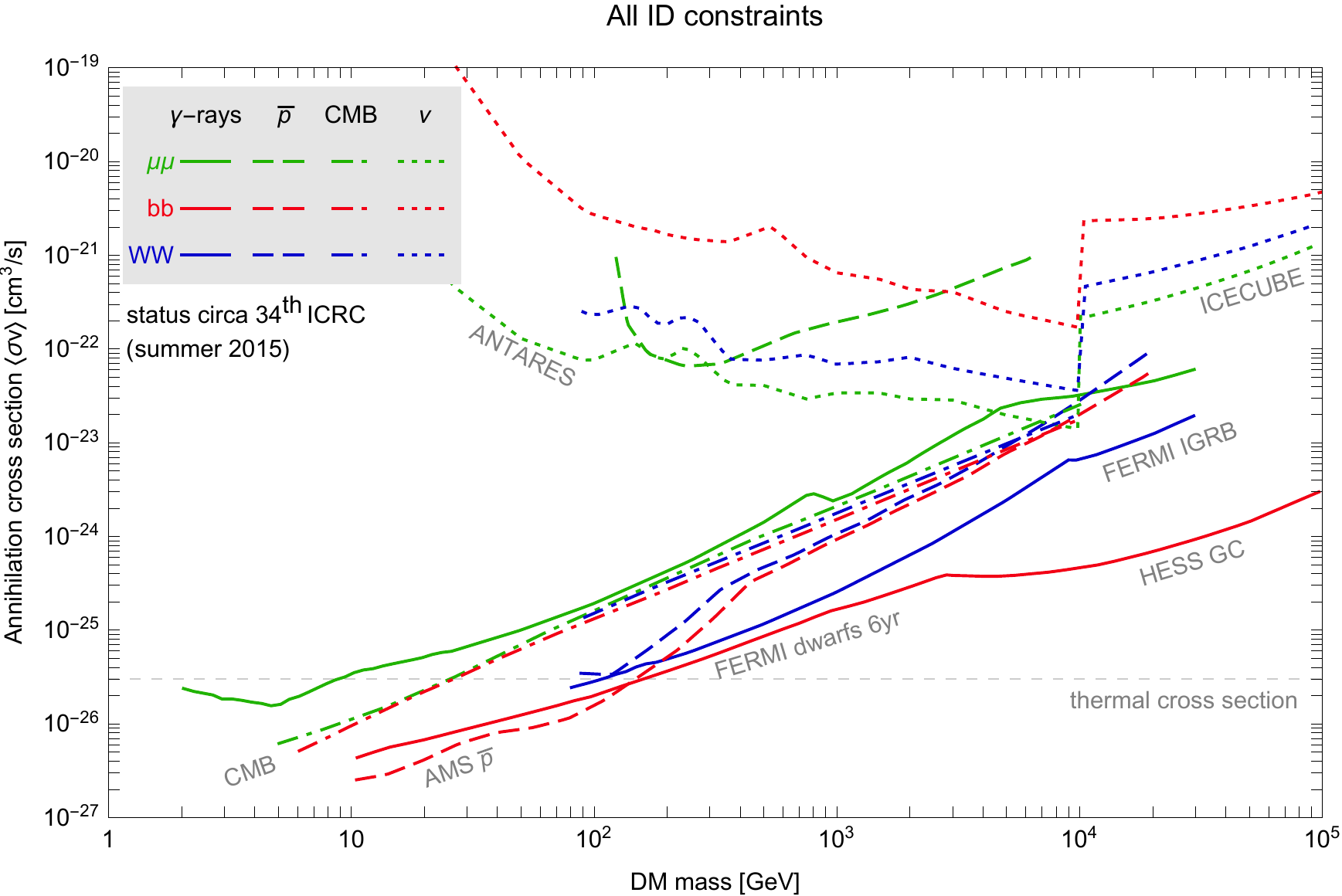} 
\caption{Summary chart of the current most stringent bounds on DM annihilation, in different channels and from different searches as discussed in the text. The corresponding references are: {\sc Ams-02} antiprotons~\cite{Giesen:2015ufa}, {\sc Fermi} dwarfs~\cite{Rico:2015nya}, CMB~\cite{Slatyer:2015jla}, {\sc Hess}-GC~\cite{Lefranc:2015vza}, {\sc Fermi} IGRB~\cite{Ackermann:2015tah}, {\sc Antares}~\cite{Adrian-Martinez:2015wey}, {\sc Icecube}~\cite{Aartsen:2014hva}. Several caveats apply: (i) `Official' bounds, i.e.~those obtained by the relevant experimental collaborations, are reported for the most part, although of course many other authors have obtained limits that may be even more constraining. (ii) When different bounds are available, `fiducial' ones are adopted. (iii) Some bounds need rescaling for the sake of a fair comparison: (iii-a) the {\sc Hess}-GC one has been rescaled to correspond to the benchmark Einasto profile defined in~\cite{Cirelli:2010xx} and used for the {\sc Ams-02} antiproton bounds; (iii-b) the {\sc Icecube} bounds have been rescaled to be comparable to the {\sc Antares} ones; (iii-c) however, a further rescaling of both neutrino bounds would be necessary to compare them with the benchmark Einasto assumption (but this is not possible as the neutrino regions of observation are not univocally defined).}
\label{fig:all_ID_constraints}
\end{center}
\end{figure}

The non-observation of high-energy neutrino fluxes from the center of the Sun, on the other hand, imposes constraints on the scattering cross section of DM particles with nuclei, the same which are relevant for DM Direct Detection (DD) (second line of fig.~\ref{fig:neutrinos}).
Remarkably, the bounds are competitive with those from the dedicated DD experiments, such as {\sc Xenon-100} or {\sc Cdms}, mostly on the spin-dependent scattering cross section but also on the spin-independent one. 
Finally, the non observation of neutrino fluxes from the center of the Earth can be used to impose constraints on the SI scattering cross section (third line of fig.~\ref{fig:neutrinos}): for some specific values of the DM mass around 50 $\div$ 60 GeV, where a resonance enhances the capture by the chemical elements constituting the Earth, they can be more stringent than the corresponding bounds from the Sun.

Before moving to another search mode, let us collect on a single plot all the constraints from indirect detection discussed so far: fig.~\ref{fig:all_ID_constraints} compares them for 3 different channels and for a variety of messengers. The important caveats discussed in the caption apply.

\section{Direct detection}
\label{sec:DD}

\begin{figure}[t]
\begin{center}
\includegraphics[width= 0.81 \columnwidth]{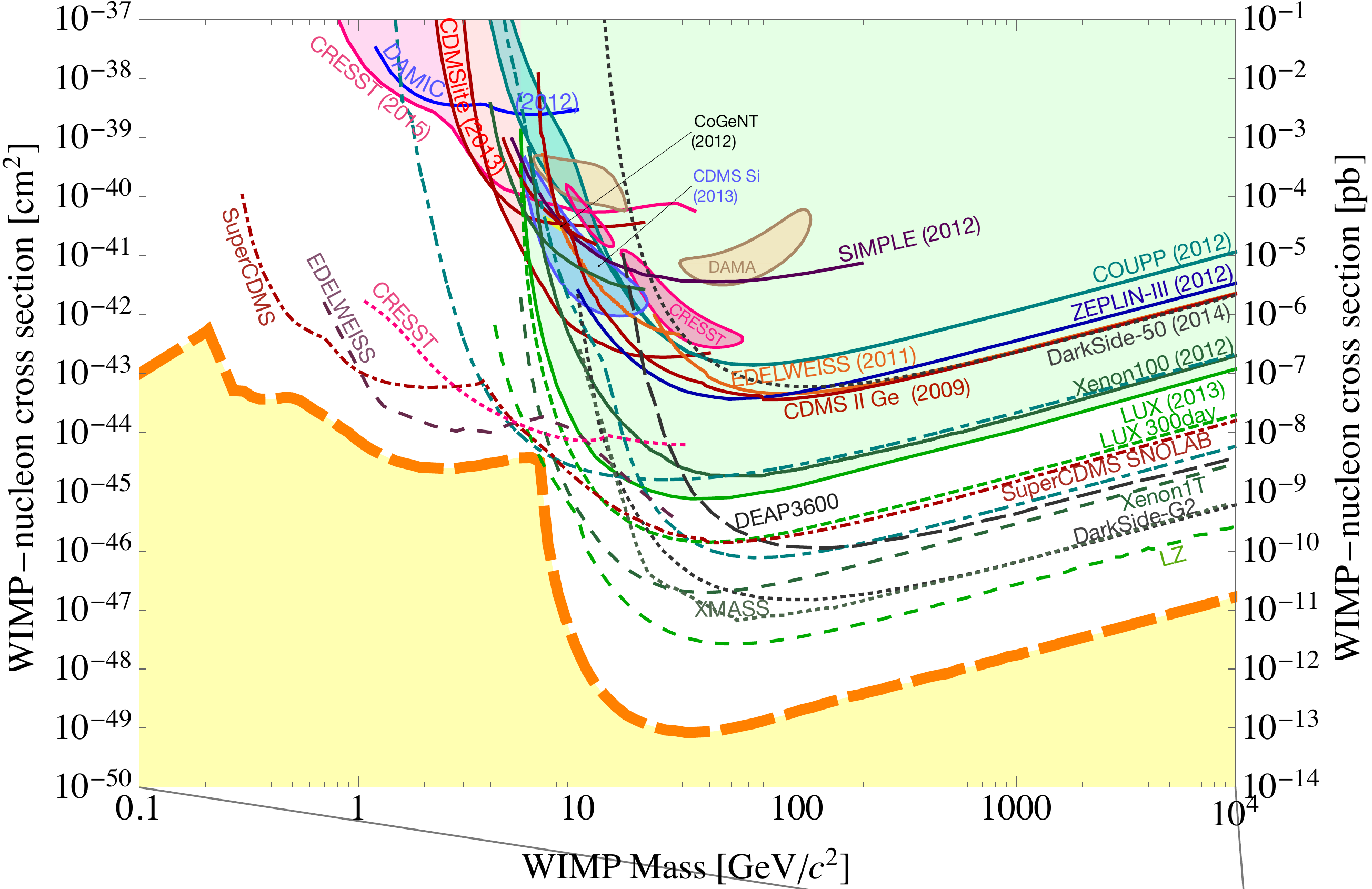} 
\caption{\label{DD} Status of Direct Detection searches and their projected sensitivity (figure from~\cite{Cushman:2013zza,FigueroaFeliciano}).}
\label{fig:DD}
\end{center}
\end{figure}

Dark Matter can also be looked for in Direct Searches, which aim at detecting, in ultra-clean and ultra-sensitive experiments, the recoil of an atom hit by a DM particle. In the most studied case, one assumes that the DM particle hits a nucleon, so that the sought-for signature consists of a nuclear recoil. But the case of electron recoil events can also be considered. Given the very low predicted DM interaction with ordinary matter (assuming that such interaction even exists!), these events are of course extremely rare, hence the experiments have to be shielded as much as possible from the `contamination' produced by ordinary cosmic rays: they are placed in deep underground caverns, shielded by km-equivalent of rock, or possibly deep under the Antarctic ice. In this sense, it is somewhat ironic that direct detection (DD) be addressed in a CR conference such as the ICRC. On the other hand, DM can be seen as yet another species of elusive radiation from space, which has to be searched for precisely by screening from the most abundant species. At ICRC 2015, a rather limited number of contributions has been devoted to this elusive radiation. 
I will limit myself to presenting the state of the art of the searches, in the form of the summary plot of fig.~\ref{DD}, and highlight the few additions discussed at the conference. The summary plot shows the DM/nucleon cross section, as a function of the DM mass, as determined by the very many experiments that have been performed so far. The upper bounds are reported (most solid lines) as well as the predicted sensitivities (most dashed lines). Some experiments (most notoriously {\sc Dama-Libra}), have been interpreted in terms of positive detection, although this is in apparent conflict with the bounds, at least under standard assumptions. The lower floor represents the region in which neutrinos, from different sources, start producing events that act as backgrounds to DM ones. 

At ICRC 2015, {\sc Damic}~\cite{DAMIC} has presented results, not reported in the summary plot, that cut an additional sliver of parameter space around $m_{\rm DM} \simeq 1 \div 3$ GeV. {\sc Xmass} has presented, with preliminary data, a bound which is currently not competitive~\cite{XMASS_results} but the collaboration aims to get at $10^{-47} {\rm cm}^2$ for a 100 GeV mass~\cite{XMASS_future}. Also, no evidence for an annual modulation in {\sc Xmass} (in the data corresponding to less than 1yr of data taking) has been reported~\cite{XMASS_modul}.

\section{Outside of the WIMP box}
\label{sec:outWIMP}

Before concluding, I will take a brief excursion outside of the WIMP framework, mentioning the contributions to ICRC 2015 that have addressed alternatives. \cite{Veberic} has started the search for hidden photon DM. Ref.~\cite{Ruchayskiy} has discussed the prospects for sterile neutrino DM, especially in the light of the X-ray line at 3.55 KeV identified in {\sc Xmm-Newton} data. In Direct Detection, ref.~\cite{Masbou} has presented bounds on electron recoil in {\sc Xenon-100}, possibly associated to Axion-like DM candidates. Ref.~\cite{Pshirkov} has considered Primordial Black Holes as DM candidates and discussed the bounds from stellar evolution. This is only a very partial glimpse in the panorama of these theory proposals and of the corresponding searches, which is extremely varied and interesting.

\section{Generic conclusions}
\label{sec:conclusions}

Dark Matter exists and discovering what it is made of is certainly one of the major open problems in particle physics and cosmology nowadays. The key to finding out the answer will probably lie in a tight collaboration among the many different disciplines involved in the quest, including in particular particle physics beyond the Standard Model and cosmic ray physics. ICRC 2015 has then of course been an ideal venue for this cross-fertilization. The potential problem, in my view, is that progress in both communities might be too slow for the needs (or the wishes) of the other community. In the recent past, there are many examples of cases in which some parts of the Dark Matter particle theory community has jumped too quickly on the interpretation of cosmic ray data, without a full understanding of the `astrophysics-related' issues and thus reaching maybe unmotivated conclusions. In the even more recent past, there are other examples of some parts of the cosmic ray community crying `Dark Matter!' too quickly, perhaps without a full control of the context.  
Given the important stakes involved, it is perhaps more worthwhile to stay focussed and work fruitfully towards the common goal.

\small

%
%
%

\vspace{1cm}

\noindent {\bf Note added:} We remind that, concerning DM ID, one can also look at pioneering works in~\cite{Khlopov}.


\begin{thebibliography}{99}
\footnotesize

\bibitem{Ade:2015xua}
  P.~A.~R.~Ade {\it et al.} [Planck Coll.],
  arXiv:1502.01589 [astro-ph.CO].
      
   \bibitem{PAMELApositrons}
O.~Adriani {\it et al.}  [PAMELA Coll.],
  Nature 458, 607-609, 2009, 
  arXiv:0810.4995.
  See also: 
O.~Adriani {\it et al.},
  Astropart.\ Phys.\  {\bf 34} (2010) 1, 
  arXiv:1001.3522.
  
\bibitem{boezio}
  M.~Boezio, 
  ``Nine Years of Cosmic Ray Investigation by the PAMELA Experiment'',
  these Proceedings, PoS ICRC {\bf 2015} (2015) 037.
 
\bibitem{HEAT}
 S.~W.~Barwick {\it et al.}  [HEAT Coll.],
  Astrophys.\ J.\  {\bf 482} (1997) L191
  [astro-ph/9703192].

\bibitem{AMS-01}
M.~Aguilar {\it et al.}  [AMS-01 Coll.],
  Phys.\ Lett.\ B {\bf 646} (2007) 145
  [astro-ph/0703154].

\bibitem{FERMIpos}
M.~Ackermann {\it et al.}  [The Fermi LAT Coll.],
  arXiv:1109.0521 [astro-ph.HE].

\bibitem{ting}
  S.~Ting, 
  ``Latest results from the Alpha Magnetic Spectrometer on the International Space Station'',
  these Proceedings, PoS ICRC {\bf 2015} (2015) 036.

\bibitem{kounine}
  A.~Kounine, 
  ``Latest Alpha Magnetic Spectrometer results : positron fraction and pbar/p ratio'',
  these Proceedings, PoS ICRC {\bf 2015} (2015) 300.
 
 \bibitem{Aguilar:2013qda}
  M.~Aguilar {\it et al.} [AMS Coll.],
  Phys.\ Rev.\ Lett.\  {\bf 110} (2013) 141102.

\bibitem{Accardo:2014lma}
  L.~Accardo {\it et al.} [AMS Coll.],
  Phys.\ Rev.\ Lett.\  {\bf 113} (2014) 121101.
  
  \bibitem{FERMIleptons}
A.~A.~Abdo {\it et al.}  [The Fermi LAT Coll.],
  Phys.\ Rev.\ Lett.\  {\bf 102} (2009) 181101,
  arXiv: 0905.0025.
    M.~Ackermann {\it et al.}  [Fermi LAT Coll.],
  Phys.\ Rev.\ D {\bf 82} (2010) 092004,
  arXiv: 1008.3999.
  
  \bibitem{duranti}
  M.~Duranti, 
  ``Precision Measurement of the (e++e-) Flux in Primary Cosmic Rays from 0.5 GeV to 1 TeV with the Alpha Magnetic Spectrometer'',
  these Proceedings, PoS ICRC {\bf 2015} (2015) 273.
  
  \bibitem{Aguilar:2014fea}
  M.~Aguilar {\it et al.} [AMS Coll.],
  Phys.\ Rev.\ Lett.\  {\bf 113} (2014) 221102.
  
\bibitem{FERMIvsAMS}
P.~Michelson, talk at the AMS-days at CERN in April 2015. Talk by R.~Bonino, TAUP 2015, Torino.
  
  \bibitem{HESSleptons}
  F.~Aharonian {\it et al.}  [H.E.S.S. Coll.],
  Phys.\ Rev.\ Lett.\  {\bf 101} (2008) 261104
  [arXiv:0811.3894].
 F.~Aharonian {\it et al.}  [H.E.S.S. Coll.],
  Astron.\ Astrophys.\  {\bf 508} (2009) 561
  [arXiv:0905.0105].

\bibitem{BorlaTridon:2011dk}
  D.~Borla Tridon {\it et al.} [MAGIC Coll.],
  arXiv:1110.4008 [astro-ph.HE].
  
\bibitem{staszak}
  D.~Staszak, ``A CR $e^-$ Spectrum with VERITAS'', these Proceedings, PoS ICRC {\bf 2015} (2015) 411.
  
\bibitem{serpico}
  P.~Serpico,``Possible physics scenarios behind CR anomalies'',
  these Proceedings, PoS ICRC {\bf 2015} 009.

\bibitem{boudaud}
  M.~Boudaud et al., 
  ``A new look at the cosmic ray $e^+$ fraction'',
  these Proceedings, PoS ICRC {\bf 2015} (2015) 1183.

\bibitem{dimauro}
  M.~Di Mauro, 
  ``Astrophysical explanation of AMS-02 electron and positron data and constraints on dark matter contribution'',
  these Proceedings, PoS ICRC {\bf 2015} (2015) 1177.

\bibitem{grimani}
  C.~Grimani, 
  ``Cosmic-ray positron measurements: on the origin of the e+ excess and limits on magnetar birthrate'',
  these Proceedings, PoS ICRC {\bf 2015} (2015) 457.
       
\bibitem{galprop}
  I.~Moskalenko, 
  ``GALPROP Code for Galactic Cosmic Ray Propagation and Associated Photon Emissions'',
  these Proceedings, PoS ICRC {\bf 2015} (2015) 492.
  A.~Strong, 
  ``Recent extensions to GALPROP'',
  these Proceedings, PoS ICRC {\bf 2015} (2015) 507.

\bibitem{Gaggero:2015mga}
  D.~Gaggero,
  ``Connections between cosmic-ray physics, gamma-ray data analysis and Dark Matter detection,''
  these Proceedings, PoS ICRC {\bf 2015} (2015) 020 
  arXiv:1509.09050 [astro-ph.HE].

\bibitem{usine}
  D.Maurin,
  ``USINE propagation code and associated tools'',
  these Proceedings, PoS ICRC {\bf 2015} 484.

\bibitem{picard}
  R.~Kissmann, O.~Reimer and A.~Strong, 
  ``Galactic cosmic ray propagation models using Picard'',
  these Proceedings, PoS ICRC {\bf 2015} (2015) 554.

\bibitem{pato}
  F.~Iocco, M.~Pato, 
  ``Mapping DM in the Milky Way'',
  these Proceedings, PoS ICRC {\bf 2015} (2015) 023.

\bibitem{Iocco:2015xga}
  F.~Iocco, M.~Pato and G.~Bertone,
  Nature Phys.\  {\bf 11} (2015) 245-248
  [arXiv:1502.03821 [astro-ph.GA]].

\bibitem{silverwood}
  H.Silverwood, 
  ``Determining the Local DM Density'',
  these Proceedings, PoS ICRC {\bf 2015} 1185.

\bibitem{Cirelli:2010xx}
  M.~Cirelli {\it et al.},
  JCAP {\bf 1103} (2011) 051
   [JCAP {\bf 1210} (2012) E01]
  [arXiv:1012.4515 [hep-ph]].
  
\bibitem{Cirelli:2008pk}
  M.~Cirelli, M.~Kadastik, M.~Raidal and A.~Strumia,
  Nucl.\ Phys.\ B {\bf 813} (2009) 1
   [Nucl.\ Phys.\ B {\bf 873} (2013) 530]
  [arXiv:0809.2409 [hep-ph]].
  
\bibitem{Slatyer:2015jla}
  T.~R.~Slatyer,
  arXiv:1506.03811 [hep-ph].
  
\bibitem{AMSpress1}
{\sc Ams} Press releases in \href{http://www.ams02.org/wp-content/uploads/2013/04/Press_AMS_en.pdf}{April 2013}, \href{http://www.ams02.org/wp-content/uploads/2014/09/AMS-PRESS-RELEASE-140918_shrink.pdf}{September 2014}, \href{http://press.web.cern.ch/press-releases/2014/09/latest-measurements-ams-experiment-unveil-new-territories-flux-cosmic-rays}{September 2014}.
  
\bibitem{Ackermann:2010ip}
  M.~Ackermann {\it et al.} [Fermi-LAT Coll.],
  Phys.\ Rev.\ D {\bf 82} (2010) 092003
  [arXiv:1008.5119].
   
\bibitem{Adriani:2015kfa}
  O.~Adriani {\it et al.},
  Astrophys.\ J.\  {\bf 811} (2015) 1,  21
  [arXiv:1509.06249 [astro-ph.HE]].

\bibitem{calet}
  H.Motz, 
  ``CALET's Sensitivity to DM and Astrophysical Sources'',
  these Proceedings, PoS ICRC {\bf 2015} 1194.
   
\bibitem{dampe}
  V.~Gallo et al., 
  ``The test results of the Silicon Tungsten Tracker of DAMPE'',
  these Proceedings, PoS ICRC {\bf 2015} (2015) 1199.
  X.~Wu, 
  ``The Silicon-Tungsten Tracker of the DAMPE Mission'',
  these Proceedings, PoS ICRC {\bf 2015} (2015) 1192.

\bibitem{CTApositrons}
  P.~Karn, 
  ``Prospects for Measuring the $e^+$ Excess with the CTA'',
  these Proceedings, PoS ICRC {\bf 2015} (2015) 799.
  
\bibitem{Adriani:2008zq}
  O.~Adriani {\it et al.},
  Phys.\ Rev.\ Lett.\  {\bf 102} (2009) 051101
  [arXiv:0810.4994 [astro-ph]].
  
  \bibitem{Adriani:2010rc}
  O.~Adriani {\it et al.} [PAMELA Coll.],
  Phys.\ Rev.\ Lett.\  {\bf 105} (2010) 121101
  [arXiv:1007.0821].
  
  \bibitem{Adriani:2012paa}
  O.~Adriani {\it et al.},
  JETP Lett.\  {\bf 96} (2013) 621
   [Pisma Zh.\ Eksp.\ Teor.\ Fiz.\  {\bf 96} (2012) 693].
  
\bibitem{amsdays}
  \href{https://indico.cern.ch/event/381134/}{{\sc Ams} days at CERN}, 15-17 april 2015, CERN.
  
\bibitem{pressrelease}
{\sc Ams} \href{http://press.web.cern.ch/press-releases/2015/04/physics-community-discuss-latest-results-ams-experiment}{Press release}, 15 April 2015.  

\bibitem{choutko}
  V.~Choutko, 
  ``Precision Measurement of the Proton Flux in Primary CRs from 1 GV to 1.8 TV with the Alpha Magnetic Spectrometer on the International Space Station'',
  these Proceedings, PoS ICRC {\bf 2015} (2015) 260.

\bibitem{Aguilar:2015ooa}
  M.~Aguilar {\it et al.} [AMS Coll.],
  Phys.\ Rev.\ Lett.\  {\bf 114} (2015) 17,  171103.

\bibitem{Adriani:2011cu}
  O.~Adriani {\it et al.} [PAMELA Coll.],
  Science {\bf 332} (2011) 69
  [arXiv:1103.4055 [astro-ph.HE]].

\bibitem{diMauro:2014zea}
  M.di Mauro, F.Donato, A.Goudelis, P.D.Serpico,
  Phys.\ Rev.\ D {\bf 90} (2014) 8,  085017, arXiv:1408.0288.

\bibitem{genolini}
  Y.~G\'enolini, 
  ``Theoretical uncertainties in extracting cosmic ray diffusion parameters: the boron to carbon ratio'',
  these Proceedings, PoS ICRC {\bf 2015} (2015) 539.

\bibitem{moskalenko}
  I.Moskalenko, 
  ``New Calculation of Secondary $\bar p$ in CRs'',
  these Proceedings, PoS ICRC {\bf 2015} 495.

\bibitem{Giesen:2015ufa}
  G.~Giesen, M.~Boudaud, Y.~G\`enolini, V.~Poulin, M.~Cirelli, P.~Salati and P.~D.~Serpico,
  JCAP {\bf 1509} (2015) 09,  023
  [arXiv:1504.04276 [astro-ph.HE]].

\bibitem{Evoli:2015vaa}
  C.~Evoli, D.~Gaggero and D.~Grasso,
  arXiv:1504.05175 [astro-ph.HE].
  
\bibitem{Kappl:2015bqa}
  R.~Kappl, A.~Reinert and M.~W.~Winkler,
  arXiv:1506.04145 [astro-ph.HE].
  
  \bibitem{boudaudpbar}
  M.~Boudaud, ``A fussy revisitation of $\bar p$ as a tool for DM searches'', PoS ICRC{\bf 2015} 1184.

\bibitem{Mambrini:2015sia}
  Y.~Mambrini, S.~Profumo and F.~S.~Queiroz,
  arXiv:1508.06635 [hep-ph].

\bibitem{Donato:1999gy}
  F.~Donato, N.~Fornengo and P.~Salati,
  Phys.\ Rev.\ D {\bf 62} (2000) 043003
  [hep-ph/9904481].

  \bibitem{vonDoetinchem}
  P.~von Doetinchem {\it et al.},
  ``Status of CR $\bar d$ searches'',
  these Proceedings, PoS ICRC {\bf 2015} (2015) 1218.
  
\bibitem{vonDoetinchem:2015zva}
  P.~von Doetinchem {\it et al.},
  ``GAPS - Dark matter search with low-energy cosmic-ray antideuterons and antiprotons,''
  these Proceedings, PoS ICRC {\bf 2015} (2015) 1219
  [arXiv:1507.02717 [astro-ph.IM]].
  
  \bibitem{BESSlimit}
H.~Fuke {\it et al.},
  Phys.\ Rev.\ Lett.\  {\bf 95} (2005) 081101,
  arXiv:astro-ph/0504361.
 
  


\bibitem{Fornengo:2011iq}
  N.~Fornengo, R.~A.~Lineros, M.~Regis and M.~Taoso,
  JCAP {\bf 1201} (2012) 005
  [arXiv:1110.4337].
  
\bibitem{bonnivard}
V.~Bonnivard {\it et al.},
  Mon.\ Not.\ Roy.\ Astron.\ Soc.\  {\bf 453} (2015) 849
  [arXiv:1504.02048].
  V.~Bonnivard {\it et al.},
  ``DM annihilation and decay factors in the MW's dSph galaxies,''
  these Proceedings, PoS ICRC {\bf 2015} (2015) 1176.
  
  \bibitem{Abramowski:2011hc}
  A.~Abramowski {\it et al.} [HESS Coll.],
  Phys.\ Rev.\ Lett.\  {\bf 106} (2011) 161301
  [arXiv:1103.3266].
  
\bibitem{Lefranc:2015vza}
  V.~Lefranc {\it et al.} [HESS Coll.],
  ``Dark matter search in the inner Galactic halo with H.E.S.S. I and H.E.S.S. II,''
  these Proceedings, PoS ICRC {\bf 2015} (2015) 1208 
  arXiv:1509.04123 [astro-ph.HE].

\bibitem{Ackermann:2012rg}
  M.~Ackermann {\it et al.} [Fermi-LAT Coll.],
  Astrophys.\ J.\  {\bf 761} (2012) 91
  [arXiv:1205.6474].

\bibitem{Aleksic:2013xea}
  J.~Aleksi\'c {\it et al.},
  JCAP {\bf 1402} (2014) 008
  [arXiv:1312.1535 [hep-ph]].
  
\bibitem{Abramowski:2014tra}
  A.~Abramowski {\it et al.} [HESS Coll.],
  Phys.\ Rev.\ D {\bf 90} (2014) 112012
  [arXiv:1410.2589].
  
\bibitem{Ackermann:2015zua}
  M.~Ackermann {\it et al.} [Fermi-LAT Coll.],
  arXiv:1503.02641 [astro-ph.HE].
  
\bibitem{Rico:2015nya}
  J.~Rico {\it et al.} [MAGIC and Fermi-LAT Coll.s],
  ``Limits to dark matter properties from a combined analysis of MAGIC and Fermi-LAT observations of dwarf satellite galaxies,''
  these Proceedings, PoS ICRC {\bf 2015} (2015)1206,   
  arXiv:1508.05827 [astro-ph.HE].

\bibitem{Drlica-Wagner:2015xua}
  A.~Drlica-Wagner {\it et al.} [Fermi-LAT and DES Coll.],
  Astrophys.\ J.\  {\bf 809} (2015) 1,  L4, arXiv:1503.02632.
 
\bibitem{Harding:2015bua}
  J.~P.~Harding {\it et al.} [HAWC Coll.],
  ``Dark Matter Annihilation and Decay Searches with the High Altitude Water Cherenkov (HAWC) Observatory,''
    these Proceedings, PoS ICRC {\bf 2015} (2015) 1213, 
  arXiv:1508.04352.
      
\bibitem{Zitzer:2015eqa}
  B.~Zitzer [VERITAS Coll.],
  ``Search for Dark Matter from Dwarf Galaxies using VERITAS,''
    these Proceedings, PoS ICRC {\bf 2015} (2015)1225, 
  arXiv:1509.01105 [astro-ph.HE].
  
\bibitem{Ackermann:2010rg}
  M.~Ackermann {\it et al.},
  JCAP {\bf 1005} (2010) 025
  arXiv:1002.2239.

\bibitem{Abramowski:2012au}
  A.Abramowski {\it et al.} [HESS Coll.],
  Astrophys.\ J.\  {\bf 750} (2012) 123
   [Astrophys.\ J.\  {\bf 783} (2014) 63]
  arXiv:1202.5494.
    
\bibitem{Pfrommer:2012mm}
  T.~Arlen {\it et al.} [VERITAS Coll.],
  Astrophys.\ J.\  {\bf 757} (2012) 123
  [arXiv:1208.0676 [astro-ph.HE]].
 
 \bibitem{Ackermann:2015fdi}
  M.~Ackermann {\it et al.} [Fermi-LAT Coll.],
  Astrophys.\ J.\  {\bf 812} (2015) 2,  159
  [arXiv:1510.00004].
    
\bibitem{Ackermann:2015tah}
  M.~Ackermann {\it et al.} [Fermi-LAT Coll.],
  JCAP {\bf 1509} (2015) 09,  008
  [arXiv:1501.05464].
    
\bibitem{Abramowski:2013ax}
  A.~Abramowski {\it et al.} [HESS Coll.],
  Phys.\ Rev.\ Lett.\  {\bf 110} (2013) 041301
  [arXiv:1301.1173].
  
\bibitem{Kieffer:2015nsa}
  M.~Kieffer et al. [HESS Coll.],
  ``Search for gamma-ray line signatures with H.E.S.S.,''
    these Proceedings, PoS ICRC {\bf 2015} (2015)1229,   
 arXiv:1509.03514.
    
 \bibitem{Ackermann:2015lka}
  M.~Ackermann {\it et al.} [Fermi-LAT Coll.],
  Phys.\ Rev.\ D {\bf 91} (2015) 12,  122002
  [arXiv:1506.00013].
  
\bibitem{FERMIclusterslines}  
B.~Anderson, S.~Zimmer, J.~Conrad, M.~Gustafsson, M.~Sánchez-Conde and R.~Caputo,
  arXiv:1511.00014.
  
  \bibitem{Ackermann:2012nb}
  M.~Ackermann {\it et al.} [Fermi-LAT Coll.],
  Astrophys.\ J.\  {\bf 747} (2012) 121
  [arXiv:1201.2691].
  
  \bibitem{Abramowski:2011hh}
  A.~Abramowski {\it et al.} [HESS Coll.],
  Astrophys.\ J.\  {\bf 735} (2011) 12
  [arXiv:1104.2548].
  
\bibitem{NietoCastano}
  D.~Nieto Castano,
  ``Hunting for dark matter subhalos among the Fermi-LAT sources with VERITAS,''
    these Proceedings, PoS ICRC {\bf 2015} (2015) 1216. 
  
\bibitem{Palacio:2015nza}
  J.~Palacio et al. [MAGIC Coll.],
  ``Constraints on the cosmic ray cluster physics from a very deep observation of the Perseus cluster with MAGIC,''
    these Proceedings, PoS ICRC {\bf 2015} (2015) 711,   
    arXiv:1509.03974.
  
\bibitem{Carr}
  J.~Carr, 
  ``Prospects for Indirect Dark Matter Searches with the Cherenkov Telescope Array (CTA)'',
  these Proceedings, PoS ICRC {\bf 2015} (2015) 1203.
  
\bibitem{Harding}
  J.~P.~Harding and B.~Dingus, 
  ``Dark Matter Annihilation and Decay Searches with the High Altitude Water Cherenkov (HAWC) Observatory'',
  these Proceedings, PoS ICRC {\bf 2015} (2015) 1227.
  
\bibitem{DiSciascio}
  G.~Di Sciascio, 
  ``Search for DM with LHAASO'',
  these Proceedings, PoS ICRC {\bf 2015} (2015) 296.



\bibitem{hooperon_history}
  L.~Goodenough and D.~Hooper,
  0910.2998 [hep-ph].
V.~Vitale {\it et al.}  [Fermi/LAT Coll.],
  0912.3828 [astro-ph.HE].
D.~Hooper and L.~Goodenough,
  Phys.\ Lett.\ B {\bf 697} (2011) 412
  [1010.2752 [hep-ph]].
A.~Morselli {\it et al.} [Fermi-LAT Coll.],
  Nuovo Cim.\ C {\bf 034N3} (2011) 311
  [1012.2292 [astro-ph.HE]].
D.~Hooper and T.~Linden,
  Phys.\ Rev.\ D {\bf 84} (2011) 123005
  [1110.0006 [astro-ph.HE]].
D.~Hooper,
  Phys.\ Dark Univ.\  {\bf 1} (2012) 1
  [1201.1303 [astro-ph.CO]].
K.~N.~Abazajian and M.~Kaplinghat,
  Phys.\ Rev.\ D {\bf 86} (2012) 083511
  [1207.6047 [astro-ph.HE]].
 D.~Hooper and T.~R.~Slatyer,
  Phys.\ Dark Univ.\  {\bf 2} (2013) 118
  [1302.6589 [astro-ph.HE]].
   C.~Gordon and O.~Macias,
  Phys.\ Rev.\ D {\bf 88} (2013) 083521
  [1306.5725 [astro-ph.HE]].
W.-C.~Huang, A.~Urbano and W.~Xue,
  1307.6862 [hep-ph].
O.~Macias and C.~Gordon,
  Phys.\ Rev.\ D {\bf 89} (2014) 6,  063515
  [1312.6671 [astro-ph.HE]].
K.~N.~Abazajian, N.~Canac, S.~Horiuchi and M.~Kaplinghat,
  Phys.\ Rev.\ D {\bf 90} (2014) 2,  023526
  [1402.4090 [astro-ph.HE]].
T.~Daylan, D.~P.~Finkbeiner, D.~Hooper, T.~Linden, S.~K.~N.~Portillo, N.~L.~Rodd and T.~R.~Slatyer,
  1402.6703 [astro-ph.HE].
 T.~Lacroix, C.~Boehm and J.~Silk,
  Phys.\ Rev.\ D {\bf 90} (2014) 4,  043508
  [1403.1987 [astro-ph.HE]].

\bibitem{Daylan:2014rsa}
T.~Daylan et al. in \cite{hooperon_history}.

\bibitem{hooperon model building}
  W.-C.~Huang, A.~Urbano and W.~Xue,
  JCAP {\bf 1404} (2014) 020
  [1310.7609].
C.~Boehm, M.~J.~Dolan, C.~McCabe, M.~Spannowsky and C.~J.~Wallace,
  JCAP {\bf 1405} (2014) 009
  [1401.6458].
A.~Hektor and L.~Marzola,
  Phys.\ Rev.\ D {\bf 90} (2014) 5,  053007
  [1403.3401].
A.~Alves, S.~Profumo, F.~S.~Queiroz and W.~Shepherd,
  Phys.\ Rev.\ D {\bf 90} (2014) 11,  115003
  [1403.5027].
A.~Berlin, D.~Hooper and S.~D.~McDermott,
  Phys.\ Rev.\ D {\bf 89} (2014) 11,  115022
  [1404.0022].
P.~Agrawal, B.~Batell, D.~Hooper and T.~Lin,
  Phys.\ Rev.\ D {\bf 90} (2014) 6,  063512
  [1404.1373].
E.~Izaguirre, G.~Krnjaic and B.~Shuve,
  Phys.\ Rev.\ D {\bf 90} (2014) 5,  055002
  [1404.2018].
C.~Boehm, M.~J.~Dolan and C.~McCabe,
  Phys.\ Rev.\ D {\bf 90} (2014) 2,  023531
  [1404.4977].
P.~Ko, W.~I.~Park and Y.~Tang,
  JCAP {\bf 1409} (2014) 013
  [1404.5257].
M.~Abdullah, A.~DiFranzo, A.~Rajaraman, T.~M.~P.~Tait, P.~Tanedo and A.~M.~Wijangco,
  Phys.\ Rev.\ D {\bf 90} (2014) 035004
  [1404.6528].
A.~Martin, J.~Shelton and J.~Unwin,
  Phys.\ Rev.\ D {\bf 90} (2014) 10,  103513
  [1405.0272].
T.~Mondal and T.~Basak,
  Phys.\ Lett.\ B {\bf 744} (2015) 208
  [1405.4877].
C.~Cheung, M.~Papucci, D.~Sanford, N.~R.~Shah and K.~M.~Zurek,
  Phys.\ Rev.\ D {\bf 90} (2014) 7,  075011
  [1406.6372].
C.~Boehm, P.~Gondolo, P.~Jean, T.~Lacroix, C.~Norman and J.~Silk,
  1406.4683.
C.~Arina, E.~Del Nobile and P.~Panci,
  Phys.\ Rev.\ Lett.\  {\bf 114} (2015) 011301
  [1406.5542].
S.~D.~McDermott,
  Phys.\ Dark Univ.\  {\bf 7-8} 12
  [1406.6408].
J.~Huang, T.~Liu, L.~T.~Wang and F.~Yu,
  Phys.\ Rev.\ D {\bf 90} (2014) 11,  115006
  [1407.0038].
C.~Balázs and T.~Li,
  Phys.\ Rev.\ D {\bf 90} (2014) 5,  055026
  [1407.0174].
N.~Okada and O.~Seto,
  Phys.\ Rev.\ D {\bf 90} (2014) 8,  083523
  [1408.2583].
K.~Ghorbani,
  JCAP {\bf 1501} (2015) 015
  [1408.4929].
N.~F.~Bell, S.~Horiuchi and I.~M.~Shoemaker,
  Phys.\ Rev.\ D {\bf 91} (2015) 2,  023505
  [1408.5142].
A.~D.~Banik and D.~Majumdar,
  Phys.\ Lett.\ B {\bf 743} (2015) 420
  [1408.5795].
M.~Cahill-Rowley, J.~Gainer, J.~Hewett and T.~Rizzo,
  JHEP {\bf 1502} (2015) 057
  [1409.1573].
J.~H.~Yu,
  Phys.\ Rev.\ D {\bf 90} (2014) 9,  095010
  [1409.3227].
J.~Guo, J.~Li, T.~Li and A.~G.~Williams,
  Phys.\ Rev.\ D {\bf 91} (2015) 9,  095003
  [1409.7864].
J.~Cao, L.~Shang, P.~Wu, J.~M.~Yang and Y.~Zhang,
  Phys.\ Rev.\ D {\bf 91} (2015) 5,  055005
  [1410.3239].
M.~Freytsis, D.~J.~Robinson and Y.~Tsai,
  Phys.\ Rev.\ D {\bf 91} (2015) 3,  035028
  [1410.3818].
M.~Heikinheimo and C.~Spethmann,
  JHEP {\bf 1412} (2014) 084
  [1410.4842].
P.~Agrawal, B.~Batell, P.~J.~Fox and R.~Harnik,
  JCAP {\bf 1505} (2015) 011
  [1411.2592].
D.~Hooper,
  Phys.\ Rev.\ D {\bf 91} (2015) 035025
  [1411.4079].
K.~Cheung, W.~C.~Huang and Y.~L.~S.~Tsai,
  JCAP {\bf 1505}, no. 05, 053 (2015)
  [1411.2619].
  K.~Ghorbani and H.~Ghorbani,
  arXiv:1501.00206.
D.~G.~Cerdeno, M.~Peiro and S.~Robles,
  Phys.\ Rev.\ D {\bf 91} (2015) 12,  123530
  [1501.01296].
M.~Kaplinghat, T.~Linden and H.~B.~Yu,
  Phys.\ Rev.\ Lett.\  {\bf 114} (2015) 21,  211303
  [1501.03507].
A.~Achterberg, S.~Amoroso, S.~Caron, L.~Hendriks, R.~Ruiz de Austri and C.~Weniger,
  JCAP {\bf 1508} (2015) 08,  006
  [arXiv:1502.05703 [hep-ph]].
T.~Gherghetta, B.~von Harling, A.~D.~Medina, M.~A.~Schmidt and T.~Trott,
  Phys.\ Rev.\ D {\bf 91} (2015) 105004
  [1502.07173].
G.~Elor, N.~L.~Rodd and T.~R.~Slatyer,
  Phys.\ Rev.\ D {\bf 91} (2015) 103531
  [1503.01773].
J.~Kopp, J.~Liu and X.~P.~Wang,
  JHEP {\bf 1504} (2015) 105
  [1503.02669].
J.~M.~Cline, G.~Dupuis, Z.~Liu and W.~Xue,
  Phys.\ Rev.\ D {\bf 91} (2015) 11,  115010
  [1503.08213].
E.~C.~F.~S.~Fortes, V.~Pleitez and F.~W.~Stecker,
  1503.08220.
K.~Ghorbani and H.~Ghorbani,
  Phys.\ Rev.\ D {\bf 91} (2015) 12,  123541,
1504.03610.
N.~Fonseca, L.~Necib and J.~Thaler,
  1507.08295.
K.~Freese, A.~Lopez, N.~R.~Shah and B.~Shakya,
  1509.05076.
M.~Duerr, P.~F.~Perez and J.~Smirnov,
  1510.07562.

\bibitem{Buchmueller:2015eea}
  O.~Buchmueller, S.~A.~Malik, C.~McCabe, B.~Penning,
  Phys.\ Rev.\ Lett.\  {\bf 115} (2015) 181802
  [1505.07826].

\bibitem{Porter:2015uaa}
  T.~A.~Porter {\it et al.} [Fermi-LAT Coll.],
  these Proceedings, PoS ICRC {\bf 2015} (2015) 815, 
  arXiv:1507.04688.
  
  \bibitem{TheFermi-LAT:2015kwa}
  [The Fermi-LAT Collaboration],
  arXiv:1511.02938 [astro-ph.HE].

\bibitem{astrovariations}
B.~Zhou, Y.~F.~Liang, X.~Huang, X.~Li, Y.~Z.~Fan, L.~Feng and J.~Chang,
  Phys.\ Rev.\ D {\bf 91} (2015) 12,  123010
  [arXiv:1406.6948].
F.~Calore, I.~Cholis and C.~Weniger,
  JCAP {\bf 1503} (2015) 038
  [arXiv:1409.0042].
  
\bibitem{Calore:2014nla}
  F.~Calore, I.~Cholis, C.~McCabe and C.~Weniger,
  Phys.\ Rev.\ D {\bf 91} (2015) 6,  063003
  [arXiv:1411.4647].

\bibitem{MSP}
K.~N.~Abazajian,
  JCAP {\bf 1103} (2011) 010
  [arXiv:1011.4275 [astro-ph.HE]].
  D.~Hooper, I.~Cholis, T.~Linden, J.~Siegal-Gaskins and T.~Slatyer,
  Phys.\ Rev.\ D {\bf 88} (2013) 083009
  [arXiv:1305.0830 [astro-ph.HE]].
  N.~Mirabal,
  Mon.\ Not.\ Roy.\ Astron.\ Soc.\  {\bf 436} (2013) 2461
  [arXiv:1309.3428 [astro-ph.HE]].
Q.~Yuan and B.~Zhang,
  JHEAp {\bf 3-4} (2014) 1
  [arXiv:1404.2318 [astro-ph.HE]].
J.~Petrovi\'c, P.~D.~Serpico and G.~Zaharijas,
  JCAP {\bf 1502} (2015) 02,  023
  [arXiv:1411.2980 [astro-ph.HE]].
I.~Cholis, D.~Hooper and T.~Linden,
  JCAP {\bf 1506} (2015) 06,  043
  [arXiv:1407.5625 [astro-ph.HE]].

\bibitem{Gaggero:2015nsa}
  D.~Gaggero, M.~Taoso, A.~Urbano, M.~Valli and P.~Ullio,
  arXiv:1507.06129 [astro-ph.HE].
  
\bibitem{Urbano}
  A.~Urbano,
  ``The Galactic center excess brought down-to-earth,''
  these Proceedings, PoS ICRC {\bf 2015} (2015) 909.

\bibitem{Carlson:2015ona}
  E.~Carlson, T.~Linden and S.~Profumo,
  arXiv:1510.04698 [astro-ph.HE].
  
\bibitem{Cholis:2015dea}
  I.~Cholis, C.~Evoli, F.~Calore, T.~Linden, C.~Weniger and D.~Hooper,
  arXiv:1506.05119 [astro-ph.HE].
  
\bibitem{Calore}
  F.~Calore et al.,
  ``Unveiling the nature of the Fermi GeV excess: robust characterisation and possible interpretations,''
  these Proceedings, PoS ICRC {\bf 2015} (2015) 915.
  
\bibitem{Petrovic:2014uda}
  J.~Petrovic, P.~D.~Serpico and G.~Zaharijas,
  JCAP {\bf 1410} (2014) 10,  052
  [arXiv:1405.7928 [astro-ph.HE]].
  
\bibitem{Carlson:2014cwa}
  E.~Carlson and S.~Profumo,
  Phys.\ Rev.\ D {\bf 90} (2014) 2,  023015
  [arXiv:1405.7685 [astro-ph.HE]].
    
\bibitem{Bartels:2015aea}
  R.~Bartels, S.~Krishnamurthy and C.~Weniger,
  arXiv:1506.05104 [astro-ph.HE].
  
\bibitem{Weniger}
  C.~Weniger et al.,
  ``Testing the interpretation of the Fermi Galactic center excess in terms of unresolved point sources,''
  these Proceedings, PoS ICRC {\bf 2015} (2015) 920.
  
  \bibitem{Lee:2015fea}
  S.~K.~Lee, M.~Lisanti, B.~R.~Safdi, T.~R.~Slatyer and W.~Xue,
  arXiv:1506.05124 [astro-ph.HE].
 
\bibitem{O'Leary:2015gfa}
  R.~M.~O'Leary, M.~D.~Kistler, M.~Kerr and J.~Dexter,
  arXiv:1504.02477 [astro-ph.HE].

 \bibitem{1411.7623}
D.~Gaggero, A.~Urbano, M.~Valli and P.~Ullio,
  Phys.\ Rev.\ D {\bf 91} (2015) 8,  083012
  [arXiv:1411.7623].
 
 \bibitem{Cirelli:2014lwa}
  M.Cirelli, D.Gaggero, G.Giesen, M.Taoso, A.Urbano,
  JCAP {\bf 1412} (2014) 12, 045, arXiv:1407.2173.
  
\bibitem{Gaggero}
  D.~Gaggero,
  ``Connections between cosmic-ray physics, gamma-ray data analysis and Dark Matter detection,''
  these Proceedings, PoS ICRC {\bf 2015} (2015) 020.

\bibitem{Bringmann:2014lpa}
  T.~Bringmann, M.~Vollmann and C.~Weniger,
  Phys.\ Rev.\ D {\bf 90} (2014) 12,  123001
  [arXiv:1406.6027].
 
 \bibitem{Hooper:2014ysa}
  D.~Hooper, T.~Linden and P.~Mertsch,
  JCAP {\bf 1503} (2015) 03,  021
  [arXiv:1410.1527 [astro-ph.HE]].
  
  \bibitem{Ipek:2014gua}
  S.~Ipek, D.~McKeen and A.~E.~Nelson,
  Phys.\ Rev.\ D {\bf 90} (2014) 5,  055021
  [arXiv:1404.3716 [hep-ph]].

\bibitem{wood}
 M.~Wood,
  ``DM searches with Fermi LAT in direction of dSphs,''
  these Proceedings, PoS ICRC {\bf 2015} 1226.

\bibitem{Abazajian:2015raa}
  K.~N.~Abazajian and R.~E.~Keeley,
  arXiv:1510.06424 [hep-ph].
    
  
  
    
  
  
\bibitem{DMnu}
  M.~Cirelli, N.~Fornengo, T.~Montaruli, I.~A.~Sokalski, A.~Strumia and F.~Vissani,
  Nucl.\ Phys.\ B {\bf 727} (2005) 99
   [Erratum-ibid.\ B {\bf 790} (2008) 338],
  hep-ph/0506298.
  
  \bibitem{Baratella:2013fya}
  P.~Baratella, M.~Cirelli, A.~Hektor, J.~Pata, M.~Piibeleht, A.~Strumia,
  JCAP {\bf 1403} (2014) 053
  [arXiv:1312.6408].
  
\bibitem{Blennow:2007tw}
  M.~Blennow, J.~Edsjo and T.~Ohlsson,
  JCAP {\bf 0801} (2008) 021
  [arXiv:0709.3898 [hep-ph]].
  
\bibitem{Rameez}
  M.~Rameez,
  ``Search for Dark Matter annihilations in the Sun using the completed IceCube neutrino telescope,''
  these Proceedings, PoS ICRC {\bf 2015} (2015) 1209.
  
\bibitem{Zoll}
  M.~Zoll,
  ``Improved methods for solar Dark Matter searches with the IceCube neutrino telescope'',
  these Proceedings, PoS ICRC {\bf 2015} (2015) 1099. 

\bibitem{Gleixner}
  A.~Gleixner,
  ``Improved methods for solar Dark Matter searches with the IceCube neutrino telescope,''
  these Proceedings, PoS ICRC {\bf 2015} (2015) 1110. 

\bibitem{Kunnen}
  J.~Kunnen,
  ``A search for Dark Matter in the centre of the Earth with the IceCube neutrino detector,''
  these Proceedings, PoS ICRC {\bf 2015} (2015) 1205. 

\bibitem{Ardid}
  M.~Ardid,
  ``Constraining Secluded Dark Matter models with the ANTARES neutrino telescope,''
  these Proceedings, PoS ICRC {\bf 2015} (2015) 1212. 

\bibitem{dewith}
  M.~De With,
  ``Searching for neutrinos from dark matter annihilations in (dwarf) galaxies and clusters with IceCube,''
  these Proceedings, PoS ICRC {\bf 2015} (2015) 1215. 
    
\bibitem{Aartsen:2014hva}
  M.~G.~Aartsen {\it et al.} [IceCube Coll.],
  Eur.\ Phys.\ J.\ C {\bf 75} (2015) 1,  20
  [arXiv:1406.6868].
  
\bibitem{Adrian-Martinez:2015wey}
  S.~Adrian-Martinez {\it et al.} [ANTARES Coll.],
  arXiv:1505.04866.
  
  
  
  
\bibitem{Cushman:2013zza}
  P.~Cushman {\it et al.},
  arXiv:1310.8327 [hep-ex].

\bibitem{FigueroaFeliciano}
  E.~Figueroa-Feliciano,
  ``DM Searches: Status and Prospects,''
  these Proceedings, PoS ICRC {\bf 2015} (2015) 006.
  
\bibitem{DAMIC}
  J.~de Mello Neto,
  ``The DAMIC DM experiment,''
  these Proceedings, PoS ICRC {\bf 2015} (2015) 1221.
    
\bibitem{XMASS_results}
  A.~Takeda,
  ``Results from the fiducial volume analysis of the XMASS-I dark matter data,''
  these Proceedings, PoS ICRC {\bf 2015} (2015) 1222.
K.~Martens,
  ``The XMASS Experimental Program and its Current Implementation,''
  these Proceedings, PoS ICRC {\bf 2015} (2015) 1214.

\bibitem{XMASS_future}
   K.Ichimura,
  ``XMASS 1.5, the next step of the XMASS experiment,''
  these Proceedings, PoS ICRC {\bf 2015} 1223.

\bibitem{XMASS_modul}
K.~Hiraide,
  ``Results from the annual modulation analysis of the XMASS-I dark matter data,''
  these Proceedings, PoS ICRC {\bf 2015} (2015) 1198.
  
  
\bibitem{Veberic}
D.~Veberic,
  ``Search for dark matter in the hidden-photon sector with a large spherical mirror,''
  these Proceedings, PoS ICRC {\bf 2015} (2015) 1191.

\bibitem{Ruchayskiy}
O.~Ruchayskiy,
  ``Decaying dark matter in X-rays?,''
  these Proceedings, PoS ICRC {\bf 2015} (2015) 032.

\bibitem{Masbou}
J.~Masbou,
  ``Recent results and status of the XENON program,''
  these Proceedings, PoS ICRC {\bf 2015} (2015) 1210.

\bibitem{Pshirkov}
M.~Pshirkov,
  ``Stellar evolution constrains primordial black holes as dark matter candidates,''
  these Proceedings, PoS ICRC {\bf 2015} (2015) 1186.

\bibitem{Khlopov}    
Ya.B.Zeldovich, A.A.Klypin, M.Yu.Khlopov and V.M.Chechetkin
Astrophysical bounds on the mass of heavy stable neutral leptons. Yadernaya
Fizika (1980) V. 31, PP. 1286-1294. [English translation: Sov.J.Nucl.Phys.
(1980) V.31, PP. 664-669]

    
\end{thebibliography}
\end{document}